\documentclass[twocolumn,eqsecnum,showpacs,preprintnumbers,amsmath,amssymb,floatfix]{revtex4}

\usepackage{graphicx}
\usepackage{dcolumn}
\usepackage{bm}

\begin{document}

\preprint{APS/123-QED}

\title{Optical bandedge of diluted magnetic semiconductors: \\
difference between II-VI and III-V-based DMSs
}

\author{Masao Takahashi}
 \email{taka@gen.kanagawa-it.ac.jp}
\affiliation{%
Kanagawa Institute of Technology\\
1030 Shimo-Ogino, Atsugi-shi, 243-0292, Japan
}%


\date{\today}

\begin{abstract}
Applying the dynamical coherent potential approximation
to a simple model,
we have theoretically studied the behavior of the optical bandedge in 
diluted magnetic semiconductors (DMSs).
For $A^{\rm II}_{1-x}$Mn$_{x}B^{\rm VI}$-type DMS,
the present study reveals that the linear relationship 
between exchange-spitting $\mit\Delta E_{ex} $ 
and the averaged magnetization $|x \langle S_z \rangle|$
 widely holds for different values of $x$.
The ratio, $\mit\Delta E_{ex}/x \langle S_z \rangle $,
however, depends not only the exchange strength but also the
band offset.
Furthermore, the present study reveals that in the low dilution of Ga$_{1-x}$Mn$_x$As
the optical bandedge exists not at the 
bandedge of the impurity band but near the bottom of host band.
The optical bandedge behaves as if the exchange interaction is ferromagnetic
although the antiferromagnetic exchange interaction actually operates at Mn site.
We conclude that the spin-dependent shift of the carrier states
between the impurity band and host band 
accompanying with the change of magnetization
causes
 the apparently ferromagnetic behavior of the optical bandedge
which was reported in the magnetoreflection measurement of
Ga$_{1-x}$Mn$_x$As.
 \end{abstract}

\maketitle

\section{Introduction}
 In order to study the effect of the {\it sp-d} exchange interaction 
between a carrier (an \textit{s} conduction electron or {\it p} hole) 
and localized magnetic moments ({\it d} spins) together with
 magnetic and chemical disorder in DMSs,
 we have previouly introduced a simple model for
 $A_{1-x}$Mn$_xB$-type DMSs \cite{taka99}.
In this model, 
the local potentials of nonmagnetic ($A$) ions
 in a semiconducting compound ($AB$) are 
substituted randomly, with mole fraction $(x)$, 
by the local potentials that include the exchange interaction 
between a carrier and the localized spin moment on a Mn (denoted by \textit{M}) ion.
 Thus, the potential to which a carrier is subjected at a site differs depending on 
whether the site is occupied by an \textit{A} ion or \textit{M} ion.
The Hamiltonian $H$ is given by
\begin{eqnarray}
   H  & = &  \sum_{m,n,\mu} \varepsilon_{mn}  a^{\dag} _{m\mu} a_{n\mu} \  
          + \sum_{n} u_n  , 
\label{Hamiltonian}
\end{eqnarray}
where $u_n$ is either $u_n^A$ (on the \textit{A} site) or $u_n^M$
 (on the \textit{M} site) 
depending on the ion species occupying the \(n\) site:
\begin{eqnarray}
   u_n^A &=& E_A  \sum_{\mu}
 a^{\dag} _{n\mu} a_{n\mu} \  , \\
   u_n^M &=& E_M  \sum_{\mu} 
a^{\dag}_{n\mu} a_{n\mu}  
   - I \sum_{\mu,\nu} a^{\dag}_{n\mu} \mbox{\boldmath $\sigma$}_{\mu \nu }
    \cdot {\bf S}_{n} a_{n\nu} \ .
\end{eqnarray}
Here, $a^{\dag }_{n\mu }$ and $a_{n\mu }$ are, respectively, 
the creation and annihilation operators
 for a carrier with spin $\mu$ on the \textit{n} site.
The transfer-matrix element between \textit{m} and \textit{n},
$\varepsilon _{mn}$, is assumed to be independent of the types of constitutional
atoms which occupy the \textit{m} and \textit{n} sites.
In II-VI-based DMSs of the $A^{\rm II}_{1-x}{\rm Mn}_xB^{\rm VI}$ type,
$E_A$ ($E_M$) represents a  nonmagnetic local potential on the $A^{2+}$ (Mn$^{2+}$) sites.
In III-V-based DMSs such as Ga$_{1-x}$Mn$_x$As, 
the spin-independent potential $E_M (<0) $ can be regarded as a screened
Coulomb attractive potential
between a carrier (hole) and the Mn$^{2+}$ ion (acceptor center). 
The exchange interaction between the carrier and localized spin 
 $\textbf{S}_n$ of the Mn site \textit{n}
is expressed by 
$-I  a^{\dag}_{n\mu} \mbox{\boldmath $\sigma$}_{\mu \nu } \cdot {\bf S}_n a_{n\nu} $,
where $ \mbox{\boldmath $\sigma$}_{\mu \nu }$ represents the element of
 the Pauli spin matrices. 
Throughout this article, we disregard the electron-electron, hole-hole, and/or
electron-hole interactions.

The virtual crystal approximation (VCA) is widely employed 
to describe the extended
states in II-VI-based DMSs \cite{Furdyna88}.
The VCA is a first-order
perturbation theory with respect to the \textit{sp-d} exchange interaction
and/or the band offset energy $(E_M-E_A)$.
In standard VCA, first molecular field approximation (MFA) is applied,
replacing $\textbf{S}_n$ by the thermal average $\langle \textbf{S}_n \rangle$
taken over all Mn site.
Second, the local potential $u_n$ is replaced by
configuration averaged one,
\begin{eqnarray}
 u^{VCA}_{n}  & = &   (1-x)u_n^A + x \langle u_n^M \rangle, \\
 &=& \sum_{\mu}  [(1-x)E_A + x (E_M-I \sigma^z_{\mu \mu } \langle S_z \rangle)] 
 a^{\dag}_{n\mu} a_{n\mu}  , \nonumber \\
\end{eqnarray}
for \textit{every} site,
where $S_z$ is the $z$-component of $\textbf{S}_n$. 
Note $\langle S_x \rangle = \langle S_y \rangle =0$
 because the magnetization is assumed to be along the $z$-axis.
A major advantage of the VCA is to employ
 the periodic Hamiltonian
\begin{eqnarray}
   H^{VCA}  & = &  \sum_{m,n,\mu} \varepsilon_{mn}  a^{\dag} _{m\mu} a_{n\mu} \  
          + \sum_{n} u_n^{VCA}  , 
\end{eqnarray}
instead of Eq.\ (\ref{Hamiltonian}).
In the VCA picture, therefore,
the carrier "sees" an effective potential  
$u^{VCA} = (1-x)E_A +x(E_M \mp I\langle S_z \rangle )$
at all site, where
$-(+)$ is for up- (down-) spin of the carrier. 
No spin-flip process of the carrier is considered.
The VCA leads to the exchange energy splitting
\begin{eqnarray}
   \mit\Delta E^{VCA}_{ex}  & = &  2xI \langle S_z \rangle .
\end{eqnarray}
Thus, the VCA apparently explains the 
energy splitting between 
$\sigma ^{+}$ and $\sigma ^{-}$ transition of {\it A}
 exciton in DMS \cite{Furdyna88,Haas91}
\begin{eqnarray}
\mit\Delta E = N_0 (\alpha-\beta) x \langle S_z \rangle 
\label{Splitting}
\end{eqnarray}
if we assume $2I$ to be the exchange constant
 $N_0\alpha$ for conduction electrons and $N_0\beta$ for valence electrons,
respectively. 
Furthermore, the VCA explains 
 the linear interpolation expression of the energy gap $E_g$ 
\begin{eqnarray}
   E_g(x)  &=&  E_g(0) + x V_{\rm eff}  \   , \label{Eg}
\end{eqnarray}
where $V_{\rm eff} \equiv E_g(1) - E_g(0)$.
Equation (\ref{Eg}) is experimentally observed  for a given composition and temperature 
in most of II-VI-based DMS when the atoms of the group II element are replaced by Mn
\cite{Furdyna88}.

There exist, however, some experimental facts that indicate
 that the application of the VCA is limited. 
Among the them, we have already tackled some problems:
the anomalous behavior of $E_g$ (bowing effect) in wide-gap DMSs \cite{taka99},
the enhancement of $N_0\beta$ with $x \rightarrow 0$
 observed in Cd$_{1-x}$Mn$_x$S \cite{taka01a},
and asymmetric splitting of Zeeman energy components \cite{taka01b}.
In the previous works, applying the coherent potential approximation (CPA)
to the present model,
we studied the carrier states of paramagnetic ($\langle S_z \rangle =0$)
and completely ferromagnetic ($\langle S_z \rangle =S$) cases. 
In the present work, extending the previous approach 
for the case of a finite $\langle S_z \rangle $,
we study the nature and property of carrier states in DMSs.

According to the VCA 
 the exchange splitting energy is proportional to
$x \langle S_z \rangle$ and 
the coefficient is equal to $N_0(\alpha -\beta )$
(see Eq. (\ref{Splitting})).
In Zn$_{1-x}$Mn$_x$Te \cite{Lascaray87} 
and Cd$_{1-x}$Mn$_x$Te \cite{Lascaray88},
however, it is reported that
the spin splitting energy is proportional to
$x \langle S_z \rangle$ but the coefficient
decreases with the increase in $x$.
This may be explained by
taking the higher order effect of 
the exchange interaction into account,
although imperfect treatment based on the second order 
perturbation was already done \cite{Bhattacharjee88}.

Another strong motivation of the present work is the
elucidation of the sign and amplitude of the \textit{p-d}
exchange interaction in Ga$_{1-x}$Mn$_x$As.
In the early stage of research, 
the \textit{ferro}magnetic (FM) coupling ($N_0\beta>0 $)
was reported on the basis of the polarized magnetoreflection measurement \cite{Szcytko96}.
However, the exchange interaction between \textit{p} holes and 
\textit{d} spins is experimentally proved later
 to be \textit{antiferro}magnetic (AFM) \cite{Oka98,Ando98,Szczytko99}.
 On the mechanism of the so-called carrier-induced ferromagnetism
 in Ga$_{1-x}$Mn$_x$As,
 which  has attracted much attention in recent years \cite{Ohno99},
we have already proposed a theory based on the present model \cite{taka02}. 
According to the theory, the occurrence of ferromagnetism closely relates 
with the magnetic impurity band which is formed due to the incorporation of
Mn to GaAs.
When an impurity band exists, the optically observed bandedge 
may be different from the bandedge of the impurity band.
Thus, it is highly desirable to clarify what the optical measurement has detected.
Throughout the present work, we investigate the difference in the optical bandedge
between II-VI-based and III-V-based DMSs.

This paper is organized as follows.
In Sec. II we briefly formulated the dynamical CPA
on the bases of the multiple-scattering theory.
The results and discussion for the behavior of the optically observed bandedge
and the carrier states in 
II-VI-based DMS (or the case with no magnetic impurity level)
is given in Sec. III.
Section IV is devoted for Ga$_{1-x}$Mn$_x$As.
Section V contains the conclusion.

\section{Dynamical coherent potential approach (CPA)}
\subsection{Dynamical CPA condition}
We shall confine our discussion to the so-called one-particle picture.
Hereafter we assume the carrier is \textit{p} hole,
although the result does not depend on the character of the carrier.
 A carrier moving in a DMS is subjected to disordered potentials
which arise not only from substitutional disorder 
but also from thermal fluctuation of \textit{d} spins
through the \textit{p-d} exchange interaction. 
Furthermore, when magnetization arises,
the effective potential for the carrier differs 
 according to the orientation of the carrier spin.
In the dynamical CPA \cite{taka96}, 
the disordered potential is considered in terms of the spin-dependent effective medium
 where a carrier is subject to a coherent potential,
 $\Sigma_{\uparrow}$ or $\Sigma_{\downarrow}$,
according to the orientation of its spin.
The coherent potential $\Sigma_{\uparrow}$ ($\Sigma_{\downarrow}$)
 is an energy ($\omega$)-dependent complex potential.
 Then, a carrier moving in this effective medium 
 is described by the unperturbed Hamiltonian $K$:
\begin{eqnarray}
    K & = &  \sum_{k\mu}(\varepsilon_k + \Sigma_{\mu})
                a\sp{\dag} _{k\mu} a_{k\mu}  \    .
\end{eqnarray}
Thus, the perturbation term $V (= H - K)$ is written 
 as a sum over each lattice site: 
\begin{eqnarray}
   V  & = &   \sum_{n}v_n \  ,   \ 
\end{eqnarray}
where $v_n$ is
 either $v_n^A$ or $v_n^M$, depending on the ion species occupying the $n$ site:
\begin{eqnarray}
   v_n^A &=& \sum_{\mu} (E_A-\Sigma _{\mu }) a^{\dag} _{n\mu} a_{n\mu}  \  , \\
   v_n^M &=& \sum_{\mu}( E_M -\Sigma _{\mu }) a^{\dag} _{n\mu} a_{n\mu}  \  
   - I \sum_{\mu,\nu} a^{\dag}_{n\mu} \mbox{\boldmath $\sigma$}_{\mu \nu }
    \cdot {\bf S}_{n} a_{n\nu} \ . \nonumber \\
\end{eqnarray}

Next, using the reference Green's function $P$ given by
\begin{eqnarray}
   P(\omega) & = & \frac{1}{\omega - K }  \    ,  \ 
\end{eqnarray}
 we define the matrix $t^A$ which represents the multiple scattering of
carriers due to the $A$ ion potential 
embedded in the effective medium by

\begin{eqnarray}
   t^A_n & = & v^A_n  [1-P v^A_n]^{-1}      \ , 
\label{tmatrixA}
\end{eqnarray}
and the matrix $t^M$ which represents the multiple scattering of
carriers due to the $M$ ion potential 
embedded in the effective medium by
\begin{eqnarray}
   t^M_n & = & v^M_n  [1-P v^M_n]^{-1}      \ . 
\label{tmatixM}
\end{eqnarray}
Note that $K$, and thus $P$, includes no localized spin operator,
 and that $t^A_n$ ($t^M_n$) represents the complete scattering 
 associated with the isolated potential $v^A_n$ ($v^M_n$) in the effective medium.  
 According to the multiple-scattering theory \cite{Gonis92, Ehrenreich76},
the total scattering operator $T$, which is related to $G \equiv 1/(\omega -H)$ as
\begin{eqnarray}
   G & = & P + P T P      \  ,
\end{eqnarray}
is expressed as the multiple-scattering series,
\begin{eqnarray}
  T & = & \sum_n t_n + \sum_n t_n P \sum_{m \ ( \neq n ) } t_m  \nonumber \\
 && + \sum_n t_n P \sum_{m \  ( \neq n ) } t_m  P \sum_{l\  ( \neq m ) } t_l + \cdots    \  .
\end{eqnarray}
 Within the single-site approximation, the condition 
\begin{eqnarray}
 \langle t_n \rangle_{\rm av} & = &  0  \qquad \mbox{ for any site {\it n} }     \  
\end{eqnarray}
leads to $\langle T \rangle_{\rm av} \cong  0$ 
and thus $\langle G \rangle_{\rm av} \cong  P$.
Here, we express the average of $t_n$ over the disorder in the system
as $ \langle t_n \rangle_{\rm av}$.
Since the present system includes both substitutional disorder
and the thermal fluctuation of the localized spin an $M$ site,
the average of the $t$ matrix is written as 
\begin{eqnarray}
 \langle t_n \rangle_{\rm av} & = &
  (1-x) t^A_n + x \langle t^M_n \rangle   \ .
\end{eqnarray}
Here, $(1-x)$ and \textit{x} are the mole fractions of \textit{A} and \textit{M} atoms, respectively;
 $\langle t^M \rangle $ means
 the thermal average of $t^M$ over fluctuating localized spin. 
In the dynamical CPA, the coherent potential $ \Sigma _{\mu }$
is decided such that the effective scattering of a carrier at the chosen site 
embedded in the effective medium is zero on average.
Note that 
the thermal average of off-diagonal $t$-matrix elements 
 $\langle t^M_{\uparrow \downarrow }  \rangle
 = \langle t^M_{\downarrow \uparrow }  \rangle =0$
 because the magnetization is assumed to be along the $z$-axis.
Therefore, the dynamical CPA condition is given by
\begin{subequations}
\begin{eqnarray}
(1-x) t^A_{\uparrow \uparrow} + x \langle t^M_{\uparrow \uparrow} \rangle &=& 0 \ , 
\label{CPAup}  \\
(1-x) t^A_{\downarrow \downarrow} + x \langle t^M_{\downarrow \downarrow} \rangle &=& 0  \ .
\label{CPAdown}
\end{eqnarray}
\label{CPAcondition}
\end{subequations}
For simplicity, the \textit{t} matrix elements 
 in the site representation  
 $<n \mu |t|n \nu >$ ($n$ is a site index, $\mu, \nu  = \uparrow$ or $ \downarrow$)
are written as $t_{\mu \nu}$.
The explicit expressions for $t^A_{\mu \nu}$ and $t^M_{\mu \nu}$
are given in Appendix A.
It is worth noting that in the expression of $t^M_{\uparrow \uparrow} 
\ (t^M_{\downarrow \downarrow}$), the spin-flip processes are  properly
taken into account.
As a result, a single \textit{t}-matrix element $t^M_{\mu \mu }$ depends
on both $\Sigma _{\uparrow }$ and $\Sigma _{\downarrow }$.
Therefore, we solve Eqs. (\ref{CPAup}) and (\ref{CPAdown}) simultaneously.
Note that the diagonal matrix element $t^M_{\mu \mu }$ 
 involves an operator $S_z$ which takes the
quantum values of $2S+1$;
 $S_z=-S,-S+1,\cdots S$. 
Thus, the thermal average over the fluctuating localized spin 
 is taken as 
\begin{equation}
\langle t^M_{\mu \mu}\rangle
 = \sum_{S_z=-S}^S t^M_{\mu \mu}(S_z)
{\rm exp} \left(\frac{hS_z}{k_BT} \right)
/\sum_{S_z=-S}^S {\rm exp} \left(\frac{hS_z}{k_BT}\right) \ ,
\label{THREM}
\end{equation}
where $h$ denotes the effective field felt by the localized spins. 
Since there is a one-to-one correspondence between
 $\langle S_z \rangle$ and the parameter
 $\lambda \equiv h/k_BT$, 
we can describe the carrier states in terms of 
$\langle S_z \rangle$ instead of $\lambda$.
In this work, we treat the localized 
 spins classically for simplicity; 
the actual calculations were performed for $S=400$.

\subsection{DOS and local DOS}
Throughout this work, we assume the model density of states 
of the semicircular form with a half-bandwidth $\Delta$, 
\begin{eqnarray}
 \rho (\varepsilon) 
  & = & \frac{2}{\pi\Delta} 
            \sqrt{1- \left(\frac{\varepsilon}{\Delta}\right)^2 } \ ,
\label{rho}
\end{eqnarray}
as an undisturbed density of states. 
Then, the density of states with $\mu$ spin, $D_{\mu}(\omega)$, is calculated by
\begin{eqnarray} 
D_{\mu}(\omega) &=&  -\frac{1}{\pi} {\rm Im} \int^{\Delta }_{-\Delta }
d\varepsilon \frac{\rho (\varepsilon  )}{\omega -\varepsilon -\Sigma _{\mu }(\omega )} \ .
\label{Dens}
\end{eqnarray}
for the $\Sigma _{\mu }$ determined by the CPA.
In all of the present numerical results, we have numerically verified 
\begin{eqnarray}
\int ^{\infty}_{-\infty} D_{\uparrow }(\omega ) d \omega 
=  \int ^{\infty}_{-\infty} D_{\downarrow }(\omega ) d \omega 
= 1 \ .
\label{norm1}
\end{eqnarray} 

The species-resolved DOS shall help us to understand the nature of the carrier states.
We calculate the \textit{A-} and \textit{M}-site components of the DOS, 
$(1-x)D^A_{\mu }(\omega )$ and $xD^M_{\mu }(\omega )$
($\mu =\uparrow $ or $\downarrow $),
where $D^A_{\mu }(\omega )$ [$D^M_{\mu }(\omega )$] 
 represents the local DOS associated with the \textit{A} (\textit{M}) site
(see Appendix B). 
Note that
\begin{eqnarray}
D_{\mu }(\omega )  &=& (1-x) D^A_{\mu }(\omega ) + x D^M_{\mu }(\omega ) \ .
\end{eqnarray} 
Since $D^A(\omega )$ and $D^M(\omega )$ are normalized,
the total number of $A$-site states and that of \textit{M}-site states are
$1-x$ and \textit{x}, respectively.  

\subsection{Optical absorption spectrum}
 $A^{\rm II}_{1-x}{\rm Mn}_x B^{\rm VI}$-type DMSs
are direct-gap semiconductors,
with the band extrema occurring at the $\Gamma $ point \cite{Furdyna88}.
Upon calculating the optical absorption spectrum,
we assume that the transition dipole moments of the $A$ and $M$ ions are same.
Under this assumption, the optical absorption spectrum is
given by the $k=0$ components of the DOS.
Since the explicit $k$ dependence of $\varepsilon _k$ is not employed in the present framework,
we assume that $k=0$ corresponds to the minimum point of the model band. 
Therefore, taking $\varepsilon _0 =-\Delta $, we define the optical absorption spectrum by: \cite{Onodera68}
\begin{eqnarray}
A_{\mu }(\omega)   &=&  -\frac{1}{\pi } {\rm Im} \frac{1}{\omega +\Delta  -\Sigma _{\mu }(\omega )}
\label{eq:a}.
\end{eqnarray}

GaAs is a direct-gap semiconductor,
whose band extrema exist at the $\Gamma $ point
\cite{Cardona64,Cohen66}.
In Ga$_{1-x}$Mn$_x$As, the bottom of the conduction band is still at $\Gamma $ point
\cite{Sanvito01},
even though an impurity band 
forms above the top of the valence band.
Therefore, we apply the optical absorption spectrum defined by Eq.\ (\ref{eq:a}).
(see later discussion in Sec. VI).

\subsection{Optical carrier spin polarization ${\cal P}(\omega )$
and spin-coupling strength $Q(\omega )$ }
In order to investigate the manner of coupling between the carrier spin and the localized spin
in DMS,
we define the optical carrier spin polarization, ${\cal P}(\omega )$, 
  by
\begin{eqnarray}
{\cal P}(\omega ) = \frac{D_{\downarrow }(\omega )-D_{\uparrow }(\omega )}
{D_{\downarrow }(\omega )+D_{\uparrow }(\omega )} \ .
\end{eqnarray} 
Further,  
we calculate the spin-coupling strength $Q(\omega )$ defined 
by (see Appendix B)
\begin{eqnarray}
Q(\omega ) & \equiv & - \left. \frac{\langle \delta (\omega -H) \ \sigma \cdot \mathbf{S} \rangle /S}
{\langle \delta (\omega -H) \rangle } \right| _{{\rm at} M \textrm{-site}} .
\end{eqnarray} 
Note that $Q(\omega )$ corresponds to $-\langle \cos \theta \rangle$,
where $\theta $ is the \textit{angle} between the carrier spin and localized spin at the
\textit{M} site.
Therefore, $Q(\omega )$ represents the strength of the spin coupling at \textit{M}-site.

\section{Results and discussion for II-VI-based DMS}
\subsection{Approximate expression for the bandedge energy shift}
Here we briefly summarize the approximate treatment
to consider the behavior of the bandedge energy,
although we can numerically solve the equation for the bandedge energy shift. 
Hereafter, we set $E_A \equiv 0$ as the origin of the energy. 
For the energy of the bottom of the band, $\omega _b$,
we assume that $\omega_b = -\Delta + \Sigma(\omega_b)$
in the dynamical CPA condition Eq. (\ref{CPAcondition}). 
Then, when $\langle S_z \rangle=S$, 
we obtain the approximate expression for the bandedge energy shift 
$\Sigma_b [=\Sigma(\omega_b)]$: \cite{taka01a}
\begin{widetext}
\begin{eqnarray}
 \frac{\Sigma_b}{\Delta} &=& \frac{1}{2}
\left\{\left(\frac{1}{2}+\frac{E_B}{\Delta}\right) 
-\sqrt{\left(\frac{1}{2}+\frac{E_B}{\Delta}\right)^2
-2x \left(\frac{E_B}{\Delta}\right)} \right\} \ ,
\label{APX}
\end{eqnarray}
\end{widetext}
with $E_B = E_M-IS$ ($E_B = E_M+IS$) 
for a carrier with $\uparrow (\downarrow )$ spin.
On the other hand,
when $\langle S_z \rangle =0$,
 we obtain the cubic equation for $\Sigma_b$: \cite{taka01b}
\begin{widetext}
\begin{eqnarray}
\Sigma_b^3 - \Sigma_b^2 (2E_M+\Delta)
 +\Sigma_b \left\{ \left(E_M-IS+\frac{\Delta }{2}\right) \left(E_M+IS+\frac{\Delta }{2}\right)
+ \frac{\Delta }{2} E_M x \right\} \nonumber \\
- \frac{\Delta }{2} x \left\{ \left(E_M-IS \right) \left(E_M+IS\right)
+ \frac{\Delta }{2} E_M  \right\} &=& 0 \ . 
\label{Shift}
\end{eqnarray}
\end{widetext}
As long as an impurity level (band) does not appears,
the energy shift of the bandedges calculated by
 using Eqs. (\ref{APX}) and (\ref{Shift})
give good approximate values,
as shown in the present work.
Hence, on the basis of the approximate expression 
we investigate the relationship between the exchange integral $N_0 \beta $
obtained by optically measurement and
the parameters $E_M$ and $IS/\Delta $ used in the present model. 
We define $N_0 \beta $ by using the exchange energy splitting
 at $\langle S_z \rangle  =S$, as:
\begin{eqnarray}
N_0\beta &=& \frac{\Sigma_b(+) - \Sigma_b(-)} { x S} \ ,
\end{eqnarray}
where $\Sigma_b (+)$ and $\Sigma_b (-)$ are the solutions of Eq. (\ref{APX})
for $E_B = E_M+IS$ and $E_B = E_M-IS$, respectively.
Therefore $N_0\beta $ is a function of $x$, and has following limiting values:
\begin{subequations}
\begin{eqnarray}
 N_0\beta & \rightarrow & \frac{2I}{\left(1+2\frac{E_M}{\Delta }\right)^2
-\left(2\frac{IS}{\Delta }\right)^2} \qquad \mbox{when $x\rightarrow 0$},
\label{x0} \nonumber \\
 \\
N_0\beta & = & 2I \hspace{3.7cm} \mbox{when $x=1$} \nonumber \\
\label{x1}
\end{eqnarray}
\end{subequations}
\begin{figure}[t]
\includegraphics[width=8.cm,clip]{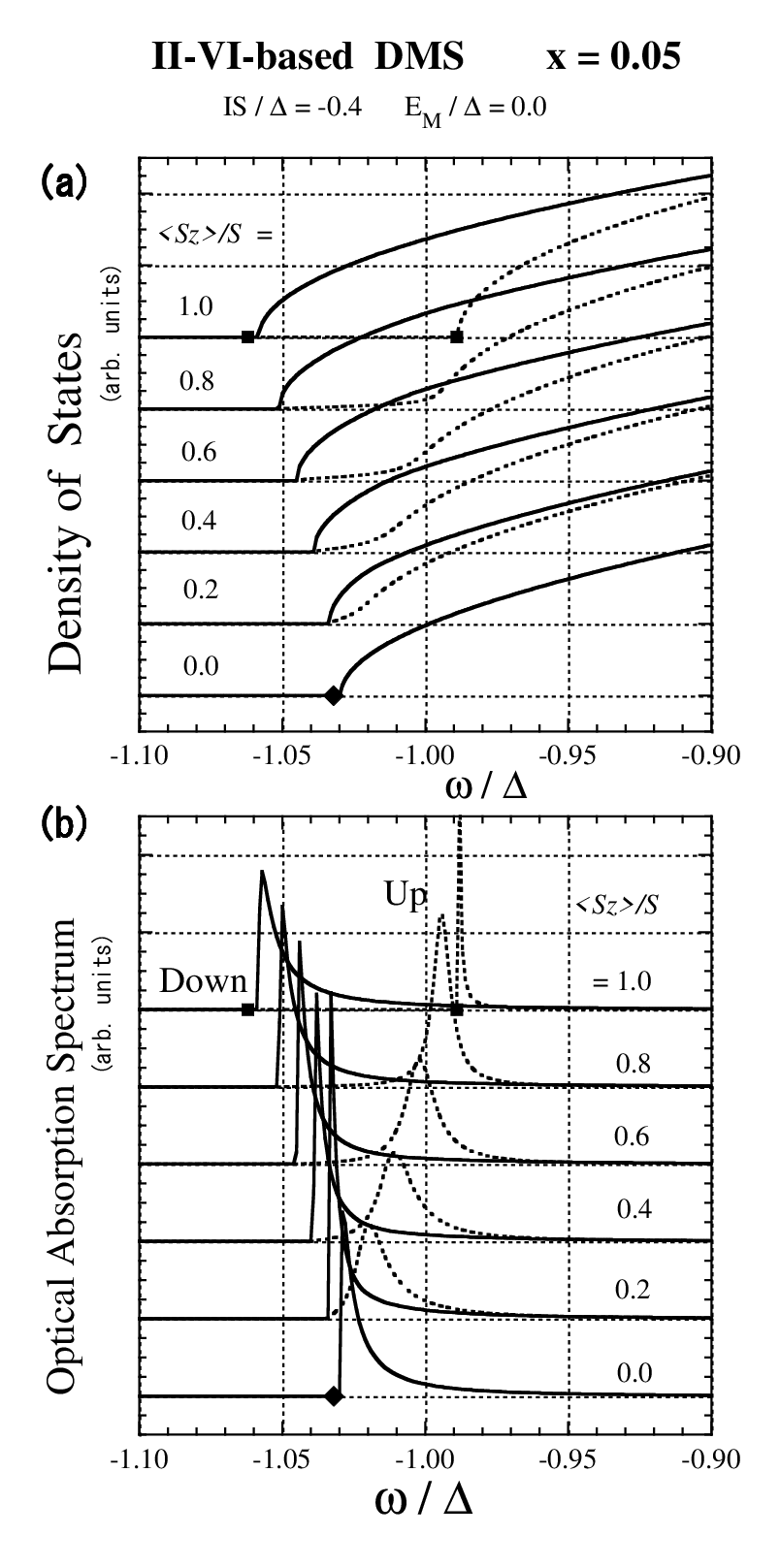}
\caption{\label{II-VIDOSb}
Results with $x=0.05$ as a function of $\omega /\Delta $
for various values of magnetization:
(a) density of states $D(\omega )$ 
(b) optical absorption band $A(\omega )$ in arbitrary unit (arb.units).
Solid line is for down-spin carrier, and dotted line is
for up-spin carrier.
The approximate values of the bandedge energy, $\omega _b/\Delta $,
 calculated using Eqs.\ (\ref{APX}) and (\ref{Shift}), 
are dotted on the line of $\langle S_z \rangle  =S$
and $\langle S_z \rangle =0$, respectively.
Note that the energy of the bottom of the model band is $\omega =-\Delta $.
}
\end{figure}

\begin{figure}[bht]
\includegraphics[width=7.cm,clip]{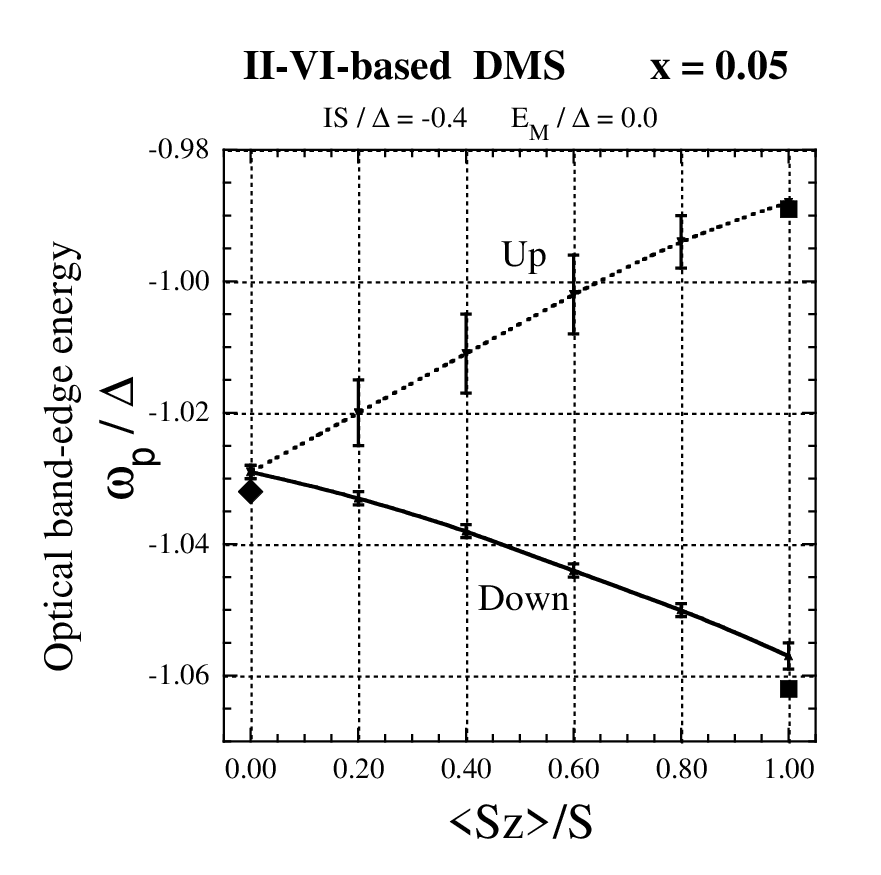}
\caption{\label{II-VISplitting}
Optical bandedge energies $\omega_p /\Delta $ as a function of 
 $\langle S_z \rangle /S$.
Solid line is for down-spin carrier, and dotted line is
for up-spin carrier.
Error bar represents a half-peak width. 
The approximate values of $\omega _b/\Delta $,
calculated by Eqs.\ (\ref{APX}) and (\ref{Shift}),
 are dotted on the line of $\langle S_z \rangle =S$ 
and $\langle S_z \rangle =0$, respectively.}
\end{figure}

\subsection{Results for $IS/\Delta =-0.4$ and $E_M/\Delta =0$}
\begin{figure}[t]
\includegraphics[width=7.cm,clip]{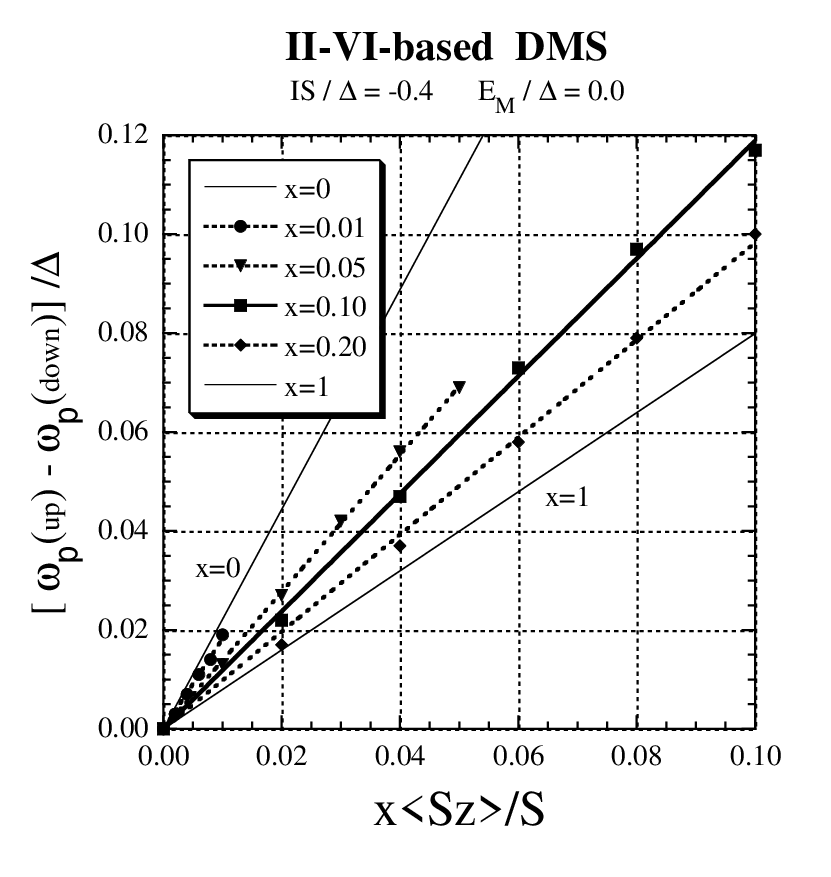}
\caption{\label{II-VIExIS04} Exchange splitting
 $[\omega_p ({\rm up} )-\omega_p ({\rm down}) ]/\Delta $ as a function of 
$x \langle S_z \rangle /S$ for various values of $x$. 
The straight lines are adjusted to the best fit the each $x$ data.
The straight lines of $x=0$ and $x=1$
exhibit the limiting cases given by Eqs.\ (\ref{x0}) and (\ref{x1}).
}
\end{figure}

 To our knowledge, no
 impurity (acceptor) level has experimentally observed in
 $A^{\rm II}_{1-x}{\rm Mn}_x B^{\rm VI}$-type DMSs.
 This implies that the fitting parameters in the present model 
should be taken as $|IS+E_M| <0.5 \Delta$ for II-VI-based DMSs.
In Figs. \ref{II-VIDOSb} $\sim $ \ref{II-VIExIS04},
 we show the results for $IS=-0.4\Delta $
and $E_M =0.0$ as a sampling case of II-VI-based DMSs.
We should note here that these parameters may be fit to describe
 a DMS like Cd$_{1-x}$Mn$_x$S
 which has the rather strong exchange interaction
among II-VI-based DMSs (see Ref. 1).

 Figure \ref{II-VIDOSb}(a) shows how the carrier band is spin-polarized 
with the development of magnetization. 
Note that even when $\langle S_z \rangle =0$,
 the band is not same as the model band. 
 Owing to the disorder of random distribution of
$M$ ion and fluctuation of localized spins,
the band has already broaden and the bottom of the band has shifted to lower side
from $\omega _b=-\Delta $.
With increase in $\langle S_z \rangle $,
the bottom of down-spin band shifts to lower-energy side
while accompanying the energy shift of bottom of the up-spin band.
The both bandedges agree with each other except the case of $\langle S_z \rangle=S$,
although the down-spin band is strongly suppressed in the band tail.
This is because the spin-flip of up-spin carrier occurs 
at the energies wherein $D_{\downarrow }(\omega )$
takes a finite value.
In magnetoabsorption and/or magnetoreflectivity spectra,
on the contrary, 
the spin splitting band is observed. 
Thus, the present result for the bandedge shift  
is very different from
the behavior expected from magneto-optical measurements.
In Fig. \ref{II-VIDOSb} (b), we show the optical absorption spectrum $A(\omega )$,
that is related to the dipole transition at $\Gamma $ point.
A  peak is found in the up- and down-optical absorption spectrum, respectively. 
Hence, we regard the energy, $\omega _p$(up) [$\omega _p$(down)], 
 at which the up (down)- spectrum 
takes a peak, as the up (down)- bandedge energy experimentally observed in optical measurement.
In Fig. \ref{II-VISplitting}, 
we display the optical bandedge energies, $\omega _p$(up) and $\omega _p$(down), 
as a function of the $\langle S_z \rangle$.
The result shows that $\omega _p$(up) and $\omega _p$(down) 
are roughly linear in $\langle S_z \rangle$. 
The behavior of the optical bandedge energy reproduces well
 an asymmetrical splitting of Zeeman energy component;
when magnetization arises
 the energy-shift pattern $\sigma ^+$ and $\sigma ^-$
transition term of the A exciton is asymmetric relative 
to the position of $\langle S_z \rangle=0$.
The asymmetric splitting of Zeeman energy component
has observed not only Cd$_{1-x}$Mn$_x$S
\cite{Gubarev90} 
but also in Cd$_{1-x}$Co$_x$Te \cite{Zielinski00}.
The half-peak width indicated by error bars in Fig. \ref{II-VISplitting}
shows that the peak of the optical spectrum with up-spin is 
broader than that of down-spin,
which may explain the reason why
the $\sigma ^{-}$ peak is broader than
$\sigma ^{+}$ peak
in the magnetoabsorption spectra \cite{Lascaray91,Twardowski79}. 

 The VCA presents the picture that a carrier in a DMS
moves freely in an uniform medium of effective potential 
$u^{VCA} = \mp xI\langle S_z \rangle $;
the depends on the orientation of the carrier spin. 
The present study reveals the feature 
of the carrier in II-VI-based DMS as below.
A carrier does not stay with the same probability on each site,
but tends to reside longer at Mn site due to the exchange interaction;
Although $x=0.05$ is assumed, the ratio of the Mn-site component of the DOS
to the total DOS $R(\omega) = 0.2 \sim 0.3$ 
 at the band tail,
as shown in Fig.\ \ref{II-VILOCALb}(c).
The carrier state at $A$-site shows similar $\langle S_z \rangle$ dependence
to that at Mn site (compare Figs.\ \ref{II-VILOCALb}(a) and (b)). 
The result of $Q(\omega ) \approx 1$ (see Fig. \ref{II-VIPQb}(b)) exhibits the
strong antiparallel spin coupling (AP-coupling) 
between carrier's spin and the localized spin 
realizes at Mn site,
and the high values of the optical spin polarization ${\cal P}(\omega )$
 (see Fig. \ref{II-VIPQb}(a))
suggests that
the carrier itinerates over the crystal while holding the effect of 
 the strong AP-coupling.
The strong spin-coupling, however, realizes  
only in the very narrow energy range of bandedge.
 The results for ${\cal P}(\omega )$
and $Q(\omega )$, shown in Fig.\ \ref{II-VIPQ}, 
suggest that 
 $Q(\omega ) = -\omega /\Delta $   
and ${\cal P}(\omega ) \approx 0 $
in the wide range of energy $\omega $ as
 $\ (-\Delta \lesssim \omega \lesssim \Delta ) $.
This is consistent with the change in $R(\omega )$.
As the energy $\omega $ increases beyond  $-\Delta $,
the probability of the carrier to reside Mn-site
decreases
so that the value of ${\cal P}(\omega )$ rapidly approaches 0. 

In Fig.\  \ref{II-VIExIS04},
the exchange-splitting energy,
$\omega _p(\text{up})-\omega _p(\text{down}) [=-\mit\Delta E_{ex}]$,
is plotted as a function of $x \langle S_z \rangle $.
The data for each $x$ are well fitted by a straight line.
The slope corresponds to $|N_0\beta |$
(or $-\mit\Delta E_{ex}/x\langle S_z \rangle=-N_0\beta $).
With the increase in $x$, the slope of the line decreases. 
It is worth noting that the straight line with $x=1$ 
agrees with that of VCA.
Thus, the present result suggests that the drastic decrease in
$|\mit\Delta E_{ex}|/ x\langle S_z \rangle$ slope
when $x$ increases.

Here, we compare the present result
 with that previously obtained by the second order perturbation.
 Considering the second-order scattering process
due to the exchange interaction,
 Bhattacharjee \cite{Bhattacharjee88}
showed that not only the coefficient  
proportional to $x \langle S_z \rangle$
but also the coefficient proportional
to $x (x\langle S_z \rangle)^{\frac{1}{2}}$
is included in the term of 
$\left(\frac{IS}{\Delta }\right)^2$.
The latter coefficient comes from that the first order correction 
is taken in to account 
in the intermediate process of the second-order scattering.
According to his result,
thus, the exchange splitting energy is not proportional to
 $x \langle S_z \rangle$.
On the other hand,
the present result suggests the proportional relationship between $\mit\Delta E_{ex}$
and  $x \langle S_z \rangle$
holds in the wide range of the parameters $IS$, $E_M$ and $x$.
The  experimental observation
in Zn$_{1-x}$Mn$_x$Te \cite{Lascaray87} and  Cd$_{1-x}$Mn$_x$Te \cite{Lascaray88}
 shows that the spin-splitting $\mit\Delta E$ as a function of $x \langle S_z \rangle$
is a straight line and that the slope
$N_0(\alpha -\beta )$ exhibits a large decrease for high $x$ values,
which seems to support the present result.

\begin{figure}[t]
\includegraphics[width=7.cm,clip]{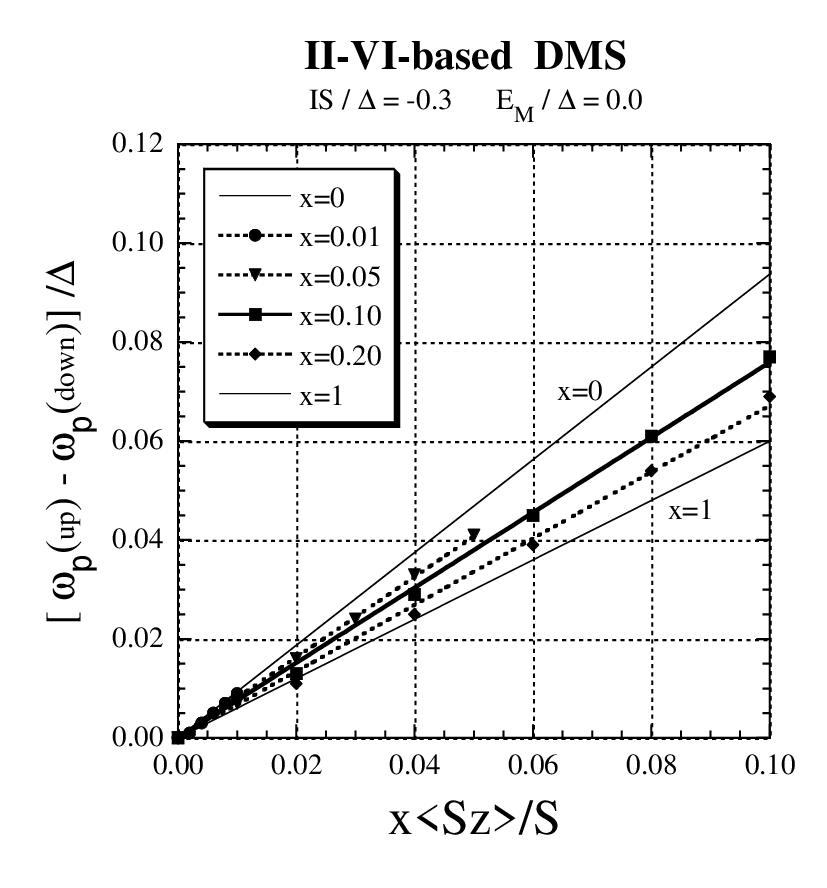}
\caption{\label{II-VIExIS03} Same as Fig.\ref{II-VIExIS04}
 but for $IS/\Delta =-0.3$ and $E_M/\Delta = 0.0$.
}
\end{figure}

\begin{figure}[t]
\includegraphics[width=7.cm,clip]{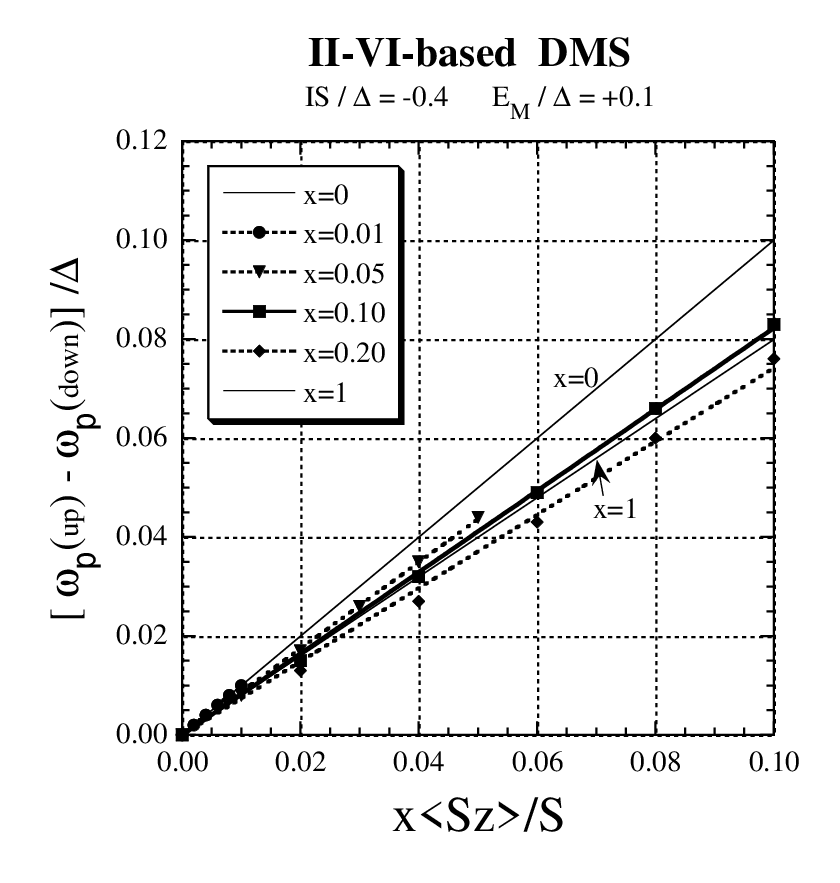}
\caption{\label{II-VIExEM01} Same as Fig.\ref{II-VIExIS04}
but for $IS/\Delta =-0.4$ and $E_M/\Delta =+0.1$.
}
\end{figure}

\subsection{Some other results for II-VI-based DMSs}
Figures \ref{II-VIExIS03} and \ \ref{II-VIExEM01}
 are the same results as Fig. \ref{II-VIExIS04} 
but for $IS =-0.3 \Delta $ and $E_M = 0.0$,
and for $IS =-0.4 \Delta $ and $E_M =+0.1 \Delta $, respectively. 
Figures indicate that the linear relationship between 
exchange splitting and $x\langle S_z \rangle$ is kept
for each case.
Furthermore, the results show that
not only the reduction in $|IS| $
but also positive $E_M$ suppress the effect of multiple scattering.
This can be understood as follows.
The repulsive interaction due to $E_M$ prevents
 the carrier to stay on Mn site longer.
Thus, positive $E_M$ substantially make the effect of the
exchange interaction weaken. 
On the other hand, negative $E_M$ assists the carrier to reside on Mn site,
resulting in apparently large $|N_0\beta| $.
Therefore, the present results request us reexamine the
VCA and the hypothesis of $N_0\beta =2I$
which have been widely accepted for II-VI-based DMSs.
We need precise information on the exchange energy, 
the bandwidth and the band offset energy 
for more quantitative comparison between
theory and experimental observation. 

\section{Results and discussion for (Ga,Mn)As}
\subsection{Model parameters for Ga$_{1-x}$Mn$_x$As}
 In this section, we discuss in detail the reason why a positive
$N_0 \beta $ is experimentally observed by magnetooptical measurement of (Ga,Mn)As.
For Ga$_{1-x}$Mn$_x$As, we take 
$\Delta = 2$ eV \cite{Shirai98,Park01}, 
 $IS= -0.4\Delta $ and $E_M= -0.3\Delta $, as the same as previous work \cite{taka02}.
These  parameters lead to an impurity level at the energy of
$E_a = -1.057\Delta $ in the dilute limit ($x \to$ 0),
which is consistent with an acceptor energy of 0.113eV $(=0.057 \Delta)$ \cite{Linnarsson97}.
With the increase in \textit{x} an impurity band forms around the
 acceptor level.
The impurity band merges into the host valence band
at $x \gtrsim 0.035$ when $\langle S_z \rangle  =0$,
while the down-spin DOSs unite 
at $x \gtrsim 0.017$ when $\langle S_z \rangle =S$.
The result roughly agrees with  
the experimental observation of impurity-band-like states 
\cite{Oka01,Oka01b}.
Also this may be related to  the insulator-metal transition

reported to occur at $x \sim 0.03$ \cite{Oiwa97,Oiwa98}.

\begin{figure}[h]
\includegraphics[width=7.cm,clip]{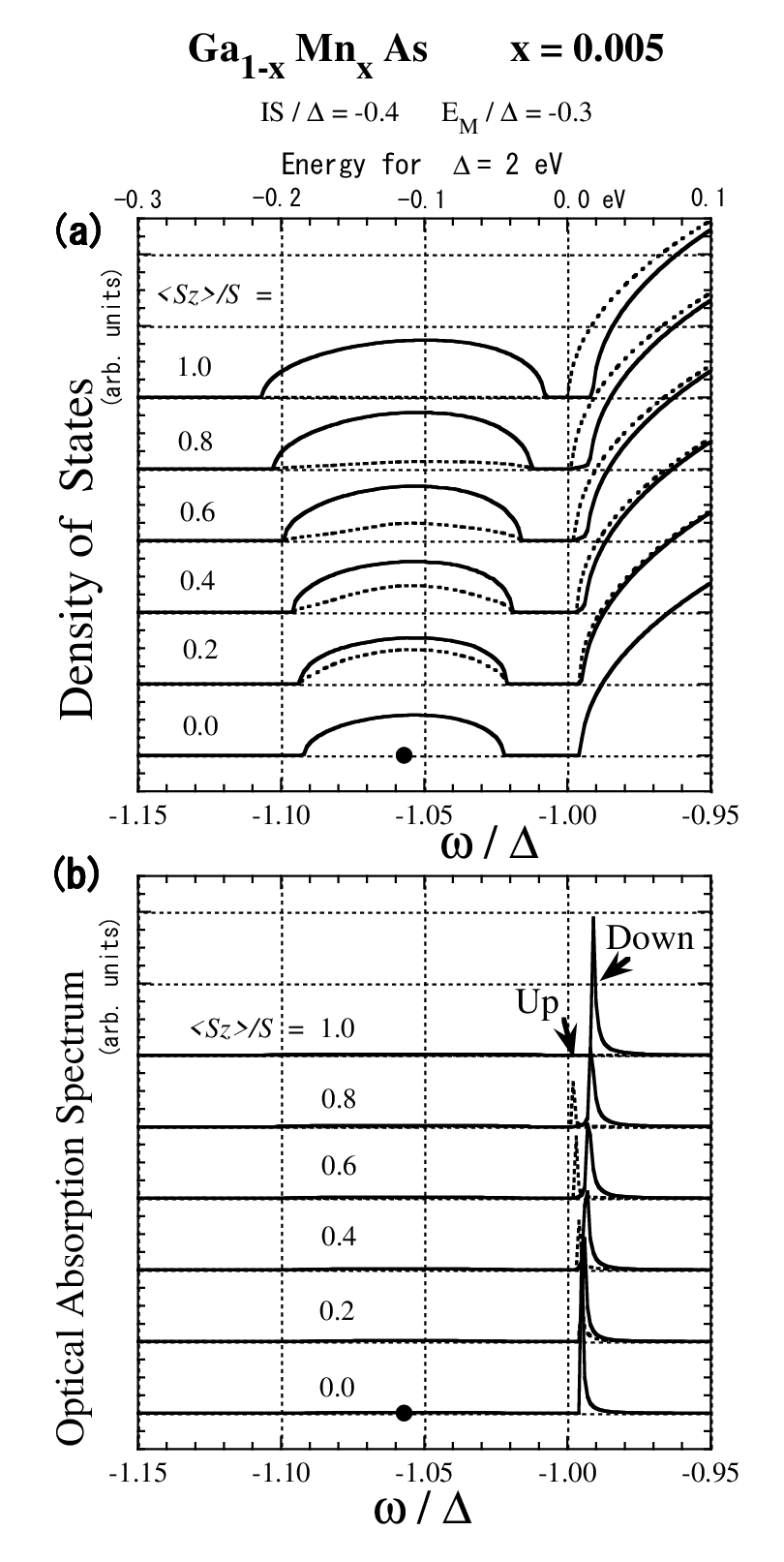}
\caption{\label{Ga05DOS} Results with $x=0.005$ as a function of $\omega /\Delta $
for various values of magnetization:
(a) density of states $D(\omega )$ 
(b) optical absorption band $A(\omega )$ in arbitrary unit (arb.units).
Solid line is for down-spin carrier, and dotted line is
for up-spin carrier.
Note that the energy of the bottom of the model band is $\omega =-\Delta $.
Above the upper horizontal axis of (a),
 energies for $\Delta =$2 eV are graduated in eV;
$\omega =-\Delta $ is taken 0 eV as the origin of the energies.
The impurity level $E_a = -1.057\Delta$ (or $ -0.057\Delta= -0.113$ eV)
is indicated by a dot on the line of $\langle S_z \rangle =0$.
}
\end{figure}

\begin{figure}[thb]
\includegraphics[width=7.cm,clip]{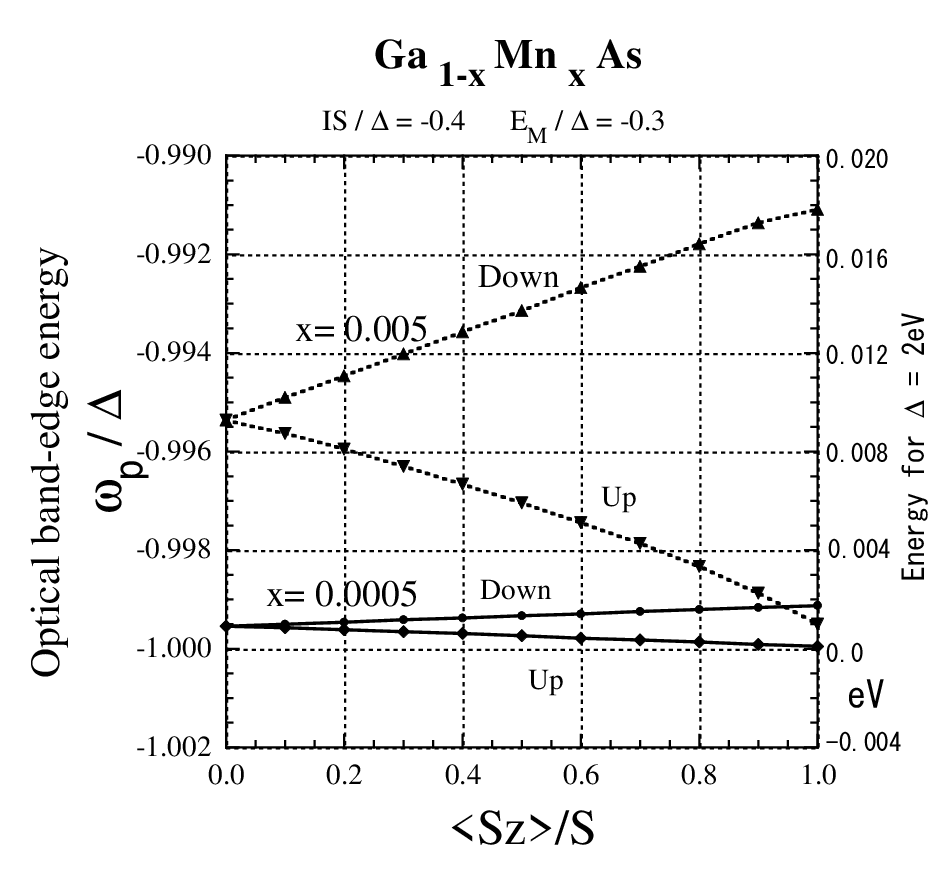}
\caption{\label{GaEdgeLow} Optical bandedge energies $\omega_p /\Delta $ 
with up- and down-spin as a function of 
$\langle S_z \rangle/S$ for $x=0.0005$ (solid line) and $x=0.005$ (dotted line). 
Beside the right vertical axis,
 energies for $\Delta =$2 eV are graduated in eV;
$\omega =-\Delta $ is taken 0 eV as the origin of the energies.
}
\end{figure}

\begin{figure}[thb]
\includegraphics[width=7.cm,clip]{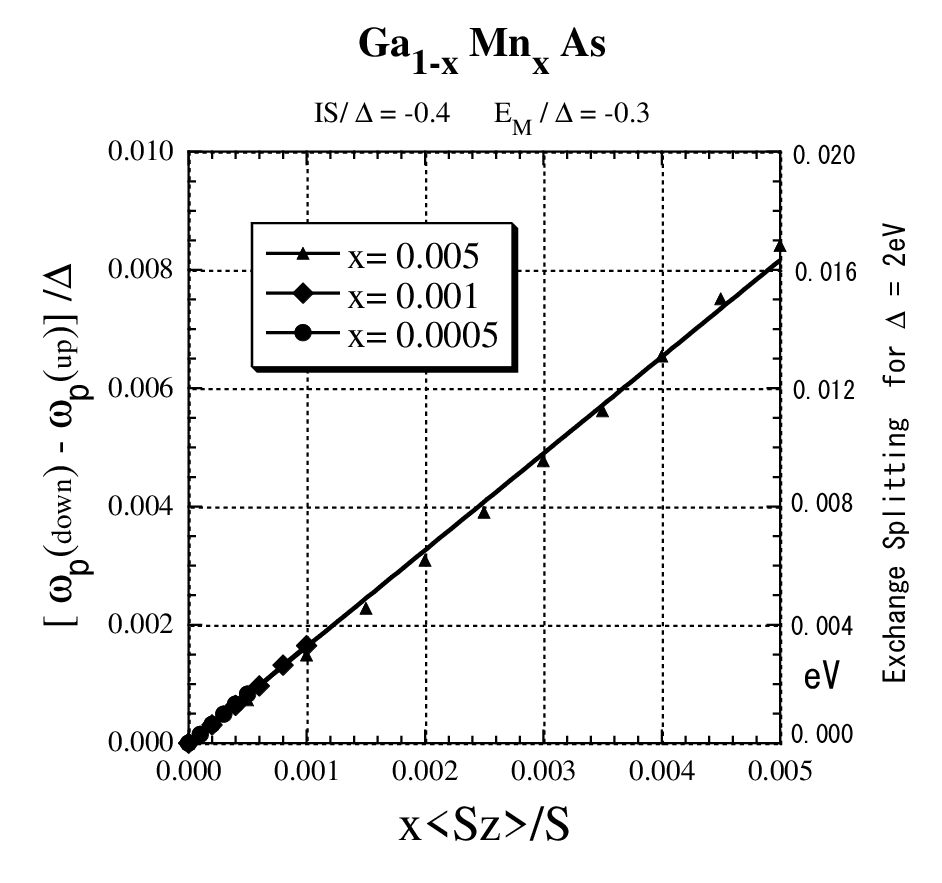}
\caption{\label{GaSplitLow} Exchange splitting
 $[\omega_p ({\rm down} )-\omega_p ({\rm up}) ]/\Delta $ as a function of 
$x \langle S_z \rangle/S$ for $x=0.0005, 0.001$, and 0.005. 
Beside the right vertical axis,
 energies for $\Delta =$2 eV are graduated in eV.
The straight line corresponds to 
 $[\omega_p ({\rm down} )-\omega_p ({\rm up})] / x \langle S_z \rangle =1.31$ eV.}
\end{figure}

\subsection{Case of low dilution }
As the typical case that $x$ is so small that an impurity band forms
separating from the host band irrespective of  $\langle S_z \rangle =S$,
we investigate the case with $x=0.005$;
the results are shown in \ref{Ga05DOS} $\sim $ \ref{Ga05LOCAL}. 
As is shown in Fig. \ref{Ga05DOS}(a),
the magnetic impurity band forms around the impurity level
 and imitates the model band.
The numerical result for the optical absorption spectrum $A(\omega )$
is shown in Fig. \ref{Ga05DOS}(b). 
We notice that 
the peak of optical absorption spectrum is almost near the bottom of the {\it host} band,
although the impurity band exists in lower energy region.
This can be explained as follows.
The states of the impurity band are composed from the states
of the wide range values in \textit{k} space.
The $k=0$ component in the impurity band states is nominal.
Since the $A(\omega )$ is related to $k=0$ state,
the optical absorption spectrum takes negligible values
in the impurity band.

As the consequence, the optical band edge, $\omega _p$, exists 
almost at the bottom of the \textit{host} band.
Here, we should stress that the energy of the peak in $A(\omega )$ of up-spin carrier
is lower than that of down-spin carrier
 (or $\omega _p(\textrm{up}) < \omega _p(\textrm{down})$),
although we have assumed the AFM exchange interaction ($IS<0$).
This implies that the direction of the shift of the optical bandedge
is opposite from that predicted by the VCA.
This may explain the reason why the FM exchange interaction
was reported on the basis of the polarized magnetoreflection measurement \cite{Szcytko96},
although the exchange interaction between \textit{p} holes and 
\textit{d} spins is AFM \cite{Oka98,Ando98,Szczytko99}.

In order to verify our picture,
we first investigate the manner of coupling
between the carrier spin and localized spins.
We show the results for ${\cal P}(\omega ) $ and  $Q(\omega ) $ in Fig.\ref{Ga05PQ}.
The result that ${\cal P}(\omega ) \approx \frac{\langle S_z \rangle}{S} $ 
and $Q(\omega ) \approx 1$ in the impurity band 
indicates that the strong antiparallel spin coupling (AP-coupling)
 between carrier spin and localized spins
occurs therein.
Near the bottom of the \textit{host} band,
however, we find that ${\cal P}(\omega ) \approx -\frac{\langle S_z \rangle}{S} $, 
suggesting the parallel spin coupling (P-coupling) therein.
On the other hand,  $Q(\omega )$ takes high values 
suggesting the weak AP-coupling near the bottom of the \textit{host} band.
Only when as $\langle S_z \rangle \gtrsim 0.95 S$,
$Q(\omega )$ takes negative values near the bottom of the host band.
Here we should notice the difference in ${\cal P}(\omega )$ and $Q(\omega )$.
The optical carrier polarization, ${\cal P}(\omega )$,
describes the spin coupling between the carrier's spin and localized spins
averaged \textit{all over the sites},
while $Q(\omega )$ represents the strength of the spin-coupling 
at \textit{M}-sites.
Hence, the result suggests that the bottom of the \textit{host} band
apparently behaves 
as if the carrier spin  \textit{ferro}magnetically couples to the localized spins 
although it may \textit{antiferro}magnetically couples to the localized spins
at \textit{M}-sites.

Next,
we investigate how different way carrier states with
up- and down-spin shift  
as the magnetization develops.
The total number of states in the impurity band is $x$,
 irrespective of $\langle S_z \rangle$. 
When $\langle S_z \rangle =0$, the carrier states with up- and down-spin  
are completely the same,
and the numbers of up- and down-spin states 
in the impurity band are $x/2$, respectively.
With increase in $\langle S_z \rangle $, as shown in Fig. \ref{Ga05DOS}(a),
the density of states with up-spin in the impurity band is suppressed,
and finally vanishes when $\langle S_z \rangle =S $.
On the other hand, the number of states with down-spin in the impurity band
 increases from $x/2$ to $x$.
The total number of states 
keeps being 1.0 per a site for each spin (up and down)
 [see Eq. (\ref{norm1})].
With the increase in $\langle S_z \rangle$, therefore,
 the carrier states with up-spin shift
from the host band to the impurity band,
whereas the carrier states with down-spin shift
from the impurity band to the host band.

To examine the mechanism in more detail,
we show the result for local DOS in Fig. \ref{Ga05LOCAL}.
Figure \ref{Ga05LOCAL}(c) indicates that rather large rate of the impurity band
is composed from \textit{M}-site states [$R(\omega ) \approx 0.5$],
in spite of small $x$ ($x=0.005$).
The total number of impurity states, 0.005,
is composed from \textit{M}-site component of 0.00246
and \textit{A}-site component of 0.00254, 
irrespective of $\langle S_z \rangle$. 
The result shown in Fig. \ref{Ga05LOCAL}(a)
indicates that the shift of the \textit{A}-site states 
occur between
 the impurity band and \textit{near the bottom of the host band}.
Figure \ref{Ga05LOCAL}(b), on the other hand, indicates that 
 the shift of the \textit{M}-site states occur
between the impurity band
and \textit{over the wide energy range of the host band}.
This can be explained as follows.
The effective local potential for carriers at \textit{M}-site
is $E_M+IS=-0.7\Delta $ for AP-coupling 
while  $E_M-IS=+0.1\Delta $ for P-coupling. 
The AP-coupling states constitute the magnetic impurity band,
whereas the P-coupling states extend over wide range of \textit{host} band 
due to the positive effective local potential. 
Thus, the up-spin carrier states at \textit{M}-site
are the AP-coupling sates in the impurity band when $\langle S_z \rangle=0$,
whereas the P-coupling states \textit{spreading over the wide-range energy of host band}
when $\langle S_z \rangle=S$.
 The carrier states with down-spin shift in the opposite way.
Hence, nominal change occurs 
\textit{near the bottom of host band}
 in \textit{M}-site component DOS, as shown in \ref{Ga05LOCAL}(b). 
The present result reveals that the shift in the bottom of the host band is
mainly ascribed to that in \textit{A}-site component DOS.
The spin dependent shift of the \textit{A}-site states near the bottom of \textit{host} band
results in the shift of optical bandedge. 
The direction of the shift is opposite from that predicted by the VCA.

In Fig. \ref{GaEdgeLow}, we show the present result for the  
optical bandedge energies $\omega_p$  
with up- and down-spins as a function of 
$\langle S_z \rangle/S$ for $x=0.0005$ and 0.005. 
The result suggests that the linear relationship between  
$\omega_p$ and $\langle S_z \rangle/S$ well holds
in such dilute case that
 the impurity band forms separate from the host band.
In Fig. \ref{GaSplitLow},
we plot the exchange splitting
 $\omega_p ({\rm down} )-\omega_p ({\rm up})\ [= +\mit\Delta E_{ex}] $ as a function of 
$x \langle S_z \rangle/S$ for $x=0.0005, 0.001$, and $x=0.005$. 
The data are well fitted by a straight line.
The $N_0\beta $ with $S=5/2$ deduced from the slop of the straight line is $+1.31$ eV.
 Szczytko \textit{et. al} measured the exciton splitting
in Ga$_{1-x}$Mn$_x$As with $x=0.00047, 0.00027$ and 0.00022
by polarized magnetoreflection,
and showed that the data
are proportional to magnetization, to
 obtain $N_0\beta =+2.5\pm 0.8$ eV
 \cite{Szcytko96}.
The agreement between the present result and experimentally obtained one
is satisfactory.
Therefore,
we conclude that the spin-dependent shift of the carrier states
between the impurity band and host band
accompanying with the change of magnetization causes
 the apparently FM behavior of the optical bandedge. 
\begin{figure}[thb]
\includegraphics[width=7.cm,clip]{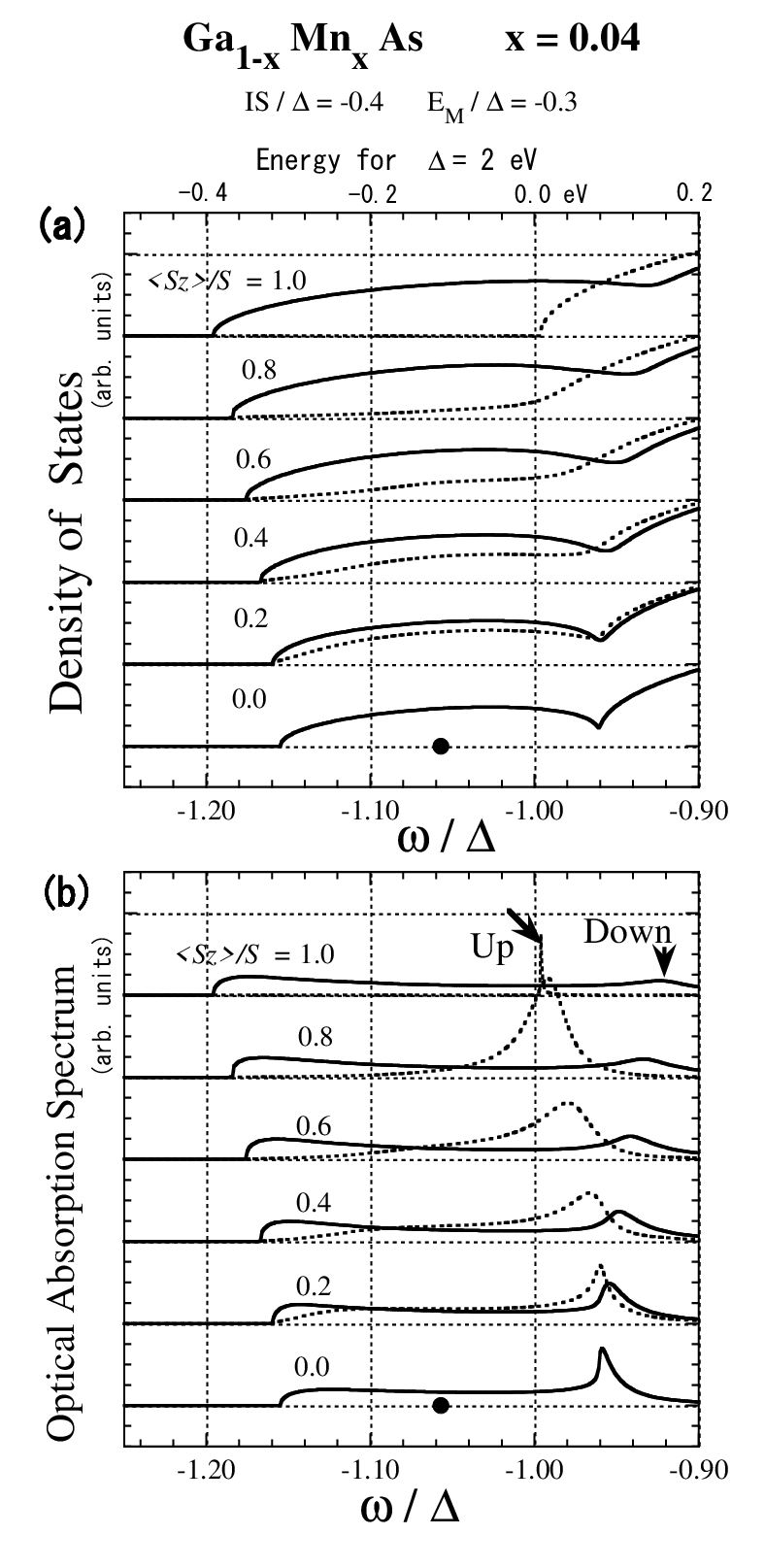}
\caption{\label{Ga4DOS} Same as Fig. \ref{Ga05DOS}, but for $x=0.04$}
\end{figure}

\begin{figure}[thb]
\includegraphics[width=7.cm,clip]{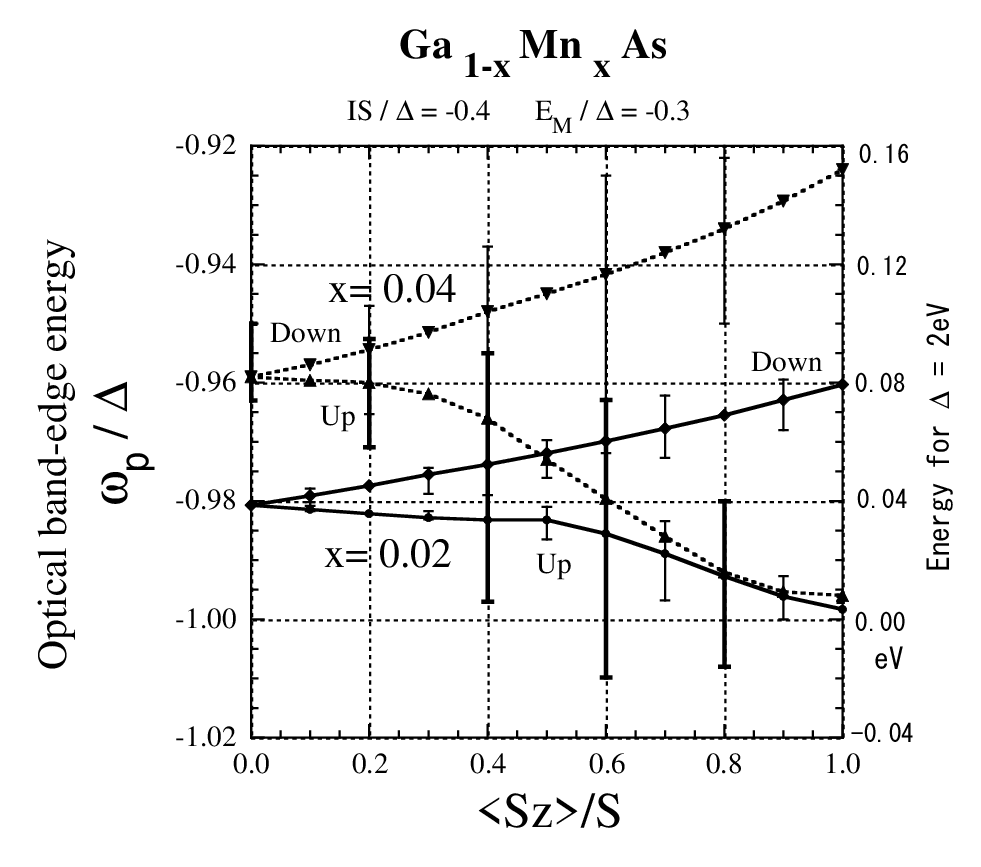}
\caption{\label{GaEdgeHigh} Optical bandedge energies $\omega_p /\Delta $ 
with up- and down-spin as a function of 
$\langle S_z \rangle/S$ for $x=0.02$ (solid line) and $x=0.04$ (dotted line). 
Error bar represents the half-peak width. 
Beside the right vertical axis,
 energies for $\Delta =$2 eV are graduated in eV;
$\omega =-\Delta $ is taken 0 eV as the origin of the energies.
}
\end{figure}

\begin{figure}[h]
\includegraphics[width=7.cm,clip]{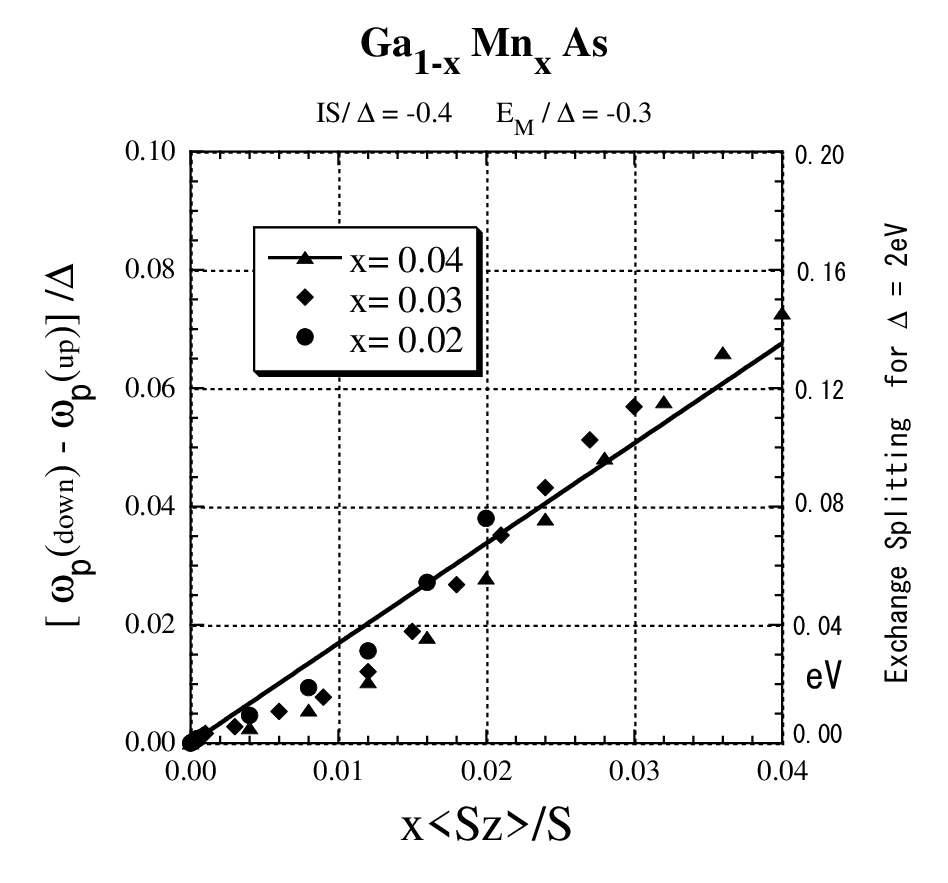}
\caption{\label{GaSplitHigh} Exchange splitting
 $[\omega_p ({\rm down} )-\omega_p ({\rm up}) ]/\Delta $ as a function of 
$x \langle S_z \rangle/S$ for $x=0.02, 0.03$, and $x=0.04$. 
Beside the right vertical axis,
 energies for $\Delta =$2 eV are graduated in eV.
The straight line corresponds to 
 $[\omega_p ({\rm down} )-\omega_p ({\rm up})] / x \langle S_z \rangle =1.36$ eV.}
\end{figure}

\subsection{Case of moderate dilution }
As the typical case that $x$ is so large that the impurity band merges
to the host band irrespective of $\langle S_z \rangle $,
we investigate the case with $x=0.04$;
the results are shown in Fig.\ref{Ga4DOS} $\sim $ \ref{Ga4LOCAL}. 
In contrast to the case of $x=0.005$,
the impurity band and host band have already united to form the band tail,
as shown in Fig.\ref{Ga4DOS}(a).
The peak of the optical absorption spectrum
is broaden due to the increase of the disorder
which is caused by the additional corporation of \textit{M} ion,
as shown in Fig.\ref{Ga4DOS}(b).
However, the peak of optical absorption spectrum
is not in the bandtail originated from the impurity band
but still exists in the energy range near the bottom of the original {\it host} band.
Thus, the direction of the shift in the optical band edge
is opposite to that predicted by the VCA,
as is the case of $x=0.005$.
The results for ${\cal P(\omega )}$ and $Q(\omega )$,
shown in Fig.\ref{Ga4PQ}, indicate that the strong AP-coupling realizes
in the bandtail,
 whereas ${\cal P(\omega )}$ shows the weak P-coupling at the energies near the optical bandedge. 
The result for the shift of the states accompanying the change
 in $\langle S_z \rangle$, deduced from Fig \ref{Ga4LOCAL},
has similar tendency with that of $x=0.0005$,
although the mergence of the impurity band makes the feature less clear.

 In Fig.\ref{GaEdgeHigh},  we 
show the optical bandedge energies, $\omega _p$(up) and $\omega _p$(down), 
as a function of 
$\langle S_z \rangle/S$ for $x=0.02$ and $x=0.04$. 
When $x=0.04$,
the absorption spectrum of up-spin is broadened.
The exchange splitting $ \mit\Delta E_{ex}$
increases monotonically with the increase in $\langle S_z \rangle/S$,
 while the linear relationship hardly holds.
In Fig. \ref{GaSplitHigh} we plot the
exchange splitting,
 $\omega_p ({\rm down} )-\omega_p ({\rm up})\ [=+\mit\Delta E_{ex}]$, as a function of 
$x \langle S_z \rangle/S$ for $x=0.02, 0.03$, and 0.04. 
The data do not fit a straight line well. 
Note that for $0.017 \lesssim x \lesssim 0.035$,
the impurity band separating from host band when $\langle S_z \rangle  =0$
unites to the host band when $\langle S_z \rangle  =S$.
The $N_0 \beta $ deduced from the slope of the straight line in Fig. \ref{GaSplitHigh} 
is +1.36 eV,
which is consistent with the experimental observation (see below).

Here we compare the present result with that of magnetoabsorption \cite{Szczytko99}.
Note that the band-gap energy of GaAs is 1.52 eV \cite{Ando98}.
In the absorption spectrum (see Fig. 1 in Ref. 12),
a rather weak structure is visible below 1.5 eV.
We regard the structure as the optical transition related to the \textit{impurity} band,
although  Szczytko \textit{et al.} disregarded it as the below-the-gap transition.
When $x$ is so large that the impurity band units the host band,
the optical absorption spectrum $A(\omega )$ takes appreciable values
at the energies in the impurity band, as shown in \ref{Ga4DOS}(b).
In the energy range above 1.5 eV, on the contrary,
the absorption band increases monotonously
 with the increase in the photon energy,
which we assign to the optical absorption related with the \textit{host} band.
Under the external magnetic field, 
the edge splits about 0.1 eV,
 which is the feature of \textit{sp-d}
exchange effect but opposite way than for Cd$_{1-x}$Mn$_x$Te.
Evaluating the spin splitting as the relative edge shift
at high energies as 1.7 eV,
they obtained $N_0(\alpha -\beta )= -2.1$ eV for $x=0.032$
and $-1.7$ eV for $x=0.042$.
Thus, they measured the energy shift at the \textit{host} band
to obtain the $N_0(\alpha -\beta )$.
The result is consistent with our view
that the optical bandedge near the bottom of original
 host band 
behaves as if the exchange interaction is FM although the
AFM exchange interaction acutely operates between the carrier and localized spins at Mn site.
The values of $N_0(\alpha -\beta )$
are comparable with $N_0(\alpha -\beta )= -(2.3 \pm 0.6)$ eV 
obtained in the very dilute cases of $x=0.00022 \sim 0.00047$
 \cite{Szcytko96}. 
 Szczytko \textit{et al.} concluded that 
the coincident of values of $N_0(\alpha -\beta )$ for different $x$ 
is accidental
because they believed the splitting inversion
is so caused by the Moss-Burstein effect that
the splitting depends on the carrier concentration.
The coincidence, however, might not be accidental
 because the spin splitting near the bottom of the \textit{host} band
was measured in both cases.
The states near the bottom of the (original) host band 
are always almost empty, 
although the carriers may enter into the impurity band or the bandtail originating from 
the impurity band.
Since the optical transition is related to the states
 near the bottom of the (original) host band,
as shown in this study,
the carrier concentration would not significantly affect 
the exchange splitting optically observed in III-V DMSs.
The impurity band and/or bandtail structure
changes with the Mn concentration $x$.
In the absorption band of Ga$_{1-x}$Mn$_x$As experimentally obtained,
the impurity-like structure of $x=0.042$ more overlaps
with the host-band-like
structure than that of $x=0.032$. 
This may be another supporter of our picture.
 
\section{Concluding remarks}
In this study,  applying the dynamical CPA
to a simple model,
 we investigated the behavior of the optical bandedge in DMSs
in a systematic way.
For $A^{\rm II}_{1-x}$Mn$_{x}B^{\rm VI}$-type DMS,
the present study reveals that the linear relationship 
between exchange-spitting $\mit\Delta E_{ex} $ 
and the averaged magnetization $|x \langle S_z \rangle|$
 widely holds for different values of $x$.
The ratio, $\mit\Delta E_{ex}/x \langle S_z \rangle $,
however, depends not only the exchange energy ($IS$) but also the
band offset ($E_M$).
The present theory can qualitatively explain the $x$ dependency
of $N_0(\alpha -\beta ) $
reported in Zn$_{1-x}$Mn$_x$Te \cite{Lascaray87} and  Cd$_{1-x}$Mn$_x$Te \cite{Lascaray88}.
For the quantitative description, however, we need more precise knowledge on the exchange energy, 
the bandwidth, and the band offset energy. 
This will be discussed in detail elsewhere.
Here, we have to indicate
that the present model
does not take into account of
 many features such as 
 multiband effects, band anisotropy, and
excitonic structure,
 which exist in real DMS's.
In the CPA, furthermore, 
the collective mode, correlation and/or clustering effect of localized spins, 
which may become significant for large $x$ region, 
are completely out of scope.
These issues remain for future study.

Regarding the \textit{p-d} exchange interaction of Ga$_{1-x}$Mn$_x$As,
there have been long controversial discussions not only on the
 amplitude but also even on the sign.
The uncertainty has made the model for the carrier-induced ferromagnetism 
 difficult to establish.
In the present paper,
we have proposed a new interpretation for the experimental result in 
 magnetooptical measurements.
A Mn$^{2+}$ ion in GaAs acts as both an acceptor and a magnetic impurity. 
Therefore, on the Mn-site, a carrier (\textit{p} hole) is subject to the
local potential which includes the \textit{p-d} exchange interaction
together with the attractive Coulomb potential.
In the dilution limit, thus, there appears an acceptor level.
With the increase in $x$, an impurity band forms around the impurity level. 
In such low dilution that the impurity band
forms separate from host band,
the optical bandedge exists not at the 
bandedge of the impurity band but near the bandedge of host band,
as shown in this study.
The optical bandedge behaves as if the exchange interaction is ferromagnetic
although the antiferromagnetic exchange interaction actually operates at Mn site.
We concluded that the spin-dependent shift of the carrier states
between the impurity band and host band
causes the apparently ferromagnetic shift of the optical bandedge.
The conclusion is valid even when $x$ is so large that the impurity band 
unites the host band
because the optical bandedge exists near the bottom of 
the original mother band.
The present result is consistent with the theory for the
mechanism of the carrier-induced ferromagnetism in III-V based
DMS's that we have previously proposed \cite{taka02}.
%

\begin{acknowledgments}
The author is grateful to Professor K. Kubo for his
helpful discussion and comments on the present study.
 This work was supported in part by Grants-in-Aid for Scientific Research 
 Nos. 14540311 from the Ministry of Education, Culture, Sports, Science and Technology of Japan.
\end{acknowledgments}

\appendix
\section{Dynamical CPA --- $t$ matrix formalism}
\subsection{$t$ matrix elements of \textit{A} ion embedded in the effective medium}
Here we omit the site suffix.
The \textit{t} matrix elements, which represent the multiple scattering of carriers
 with $\uparrow (\downarrow)$-spin due to
the \textit{A} ion potential $E_A$ embedded in the effective medium
 $\Sigma_{\uparrow}$  $(\Sigma_{\downarrow})$, 
is given by \cite{Gonis92}
\begin{subequations}
\begin{eqnarray}
 t^A_{\uparrow \uparrow}  &=& \frac{E_A-\Sigma_{\uparrow}}{1-(E_A-\Sigma_{\uparrow})F_{\uparrow}} \ , 
\label{tAup} \\
 t^A_{\downarrow \downarrow}  &=& \frac{E_A-\Sigma_{\downarrow}}
{1-(E_A-\Sigma_{\downarrow})F_{\downarrow}} \ .
\label{tAdown}
\end{eqnarray}
\end{subequations}
Here, $F_{\mu} [\equiv F_{\mu}(\omega )] $ is the diagonal matrix element of 
a propagator $P$ with respect to the effective medium, 
$\Sigma_{\mu} [\equiv \Sigma_{\mu}(\omega )]$ ($\mu =\uparrow $ or $\downarrow $),
and is calculated by 
\begin{eqnarray}
 F_{\mu}  = \langle \mu |P|\mu \rangle
  & = &     \int_{-\Delta}^{\Delta} d \varepsilon 
        \frac{\rho (\varepsilon )}{\omega - \varepsilon -\Sigma_{\mu}(\omega )} \ ,  
\label{F}
\end{eqnarray}
where $\rho (\omega )$ is the model DOS.
For $\rho (\omega )$ given by Eq. (\ref{rho}), we obtain
\begin{eqnarray}
 F_{\mu}(\omega )\Delta 
  & = &  2 \left\{ \left( \frac{\omega -\Sigma _\mu }{\Delta }\right)
  - \sqrt{ \left( \frac{\omega -\Sigma _\mu }{\Delta }\right)^2 - 1 } \right\} \ .
\end{eqnarray}
Note that $t_{\uparrow \downarrow}^A=t_{\downarrow \uparrow}^A=0.$

\subsection{$t$ matrix elements of \textit{M} ion embedded in the effective medium}
In accordance with the definition of the \textit{t} matrix,
 Eq.\ (\ref{tmatixM}),
we have
\begin{eqnarray}
 t^M [1-Pv^M] &=& v^M \ .
\label{tmatrixel}
\end{eqnarray}
Equation (\ref{tmatrixel}) is written in the spin-matrix element expression  as
\begin{eqnarray}
 t^M_{\uparrow \uparrow} -  t^M_{\uparrow \uparrow} F_{\uparrow } v^M_{\uparrow \uparrow}
 -  t^M_{\uparrow \downarrow} F_{\downarrow} v^M_{\downarrow \uparrow} 
& = &  v^M_{\uparrow \uparrow} \ , 
\label{tupup}
\\
 t^M_{\uparrow \downarrow} -  t^M_{\uparrow \downarrow} F_{ \downarrow} v^M_{\downarrow \downarrow}
 -  t^M_{\uparrow \uparrow} F_{\uparrow} v^M_{\uparrow \downarrow} &=&  v^M_{\uparrow \downarrow}
\label{tupdn} \ .
\end{eqnarray}
Then, Eq. (\ref{tupup}) $\times (F_{\downarrow}v_{\downarrow \uparrow})^{-1} $
+ Eq. (\ref{tupdn}) $\times (1-F_{\downarrow}v_{\downarrow \downarrow})^{-1}$ 
leads to an equation including $t_{\uparrow\uparrow}$ only 
($t_{\uparrow\downarrow}$ is canceled):
\begin{widetext}
\begin{eqnarray}
 \ t^M_{\uparrow \uparrow}
 [(1-F_{\uparrow} v^M_{\uparrow \uparrow})(F_{\downarrow}v^M_{\downarrow \uparrow})^{-1}
 - F_{\uparrow}v^M_{\uparrow \downarrow}(1-F_{\downarrow} v^M_{\downarrow \downarrow})^{-1}]
=  v^M_{\uparrow \uparrow} (F_{\downarrow}v^M_{\downarrow \uparrow})^{-1} 
 + v^M_{\uparrow \downarrow} (1-F_{\downarrow}v^M_{\downarrow \downarrow})^{-1}   . \ 
\end{eqnarray}
\end{widetext}
By using the following definitions and/or symbols introduced for simplicity, 
\begin{subequations}
\begin{eqnarray}
V_{\uparrow} & \equiv  & v^M_{\uparrow \uparrow } = E_M - IS_{z}- \Sigma_{ \uparrow} \ , \\  
V_{\downarrow} & \equiv  & v^M_{\downarrow \downarrow } = E_M + IS_{z} - \Sigma_{ \downarrow} \ , \\
v^M_{\uparrow \downarrow } & = & - IS_{-} \\
v^M_{\downarrow \uparrow } & = & - IS_{+} \\
 U_{\uparrow} & \equiv  & E_M - I(S_{z}-1)- \Sigma_{ \uparrow} \ , \\  
 U_{\downarrow} & \equiv & E_M + I(S_{z}+1) - \Sigma_{ \downarrow}  \ , \\
W_{\uparrow}   &\equiv&   I^{2} S_{-} S_{+} = I^2 [S(S+1)-S_z^2-S_z] \ ,  
 \\          
W_{\downarrow} &\equiv&   I^{2} S_{+} S_{-}  = I^2 [S(S+1)-S_z^2+S_z]
 \ ,  
\end{eqnarray}
\label{V}
\end{subequations}
and recalling the commutation relationships between the components of \textbf{S}, 
\begin{eqnarray}
 S_{-}S_{z} & = & (S_z+1)S_{-} \ , \\
  (S_{+})^{-1} & = & [S(S+1)- (S_z)^2-S_z]^{-1}(S_{-}) \ ,
\end{eqnarray}
we obtain an explicit expression for $t^M_{\uparrow\uparrow}$
using no more approximations.
Other \textit{t}- matrix elements are obtained by a similar procedure.
The resulting expressions are
\begin{subequations}
\begin{eqnarray}
t^M_{\uparrow\uparrow}   & = & \frac{V_{\uparrow}
       +F_{\downarrow} (W_{\uparrow}-V_{\uparrow}U_{\downarrow})}
      {1-F_{\downarrow}U_{\downarrow}-F_{\uparrow}
       [V_{\uparrow}+F_{\downarrow}
            (W_{\uparrow}-V_{\uparrow}U_{\downarrow})] }  \ , 
\label{tup} \\
t^M_{\downarrow\downarrow}   & = & \frac{V_{\downarrow}
       +F_{\uparrow}(W_{\downarrow}-V_{\downarrow}U_{\uparrow})}
      {1-F_{\uparrow}U_{\uparrow}-F_{\downarrow}
       [V_{\downarrow}+F_{\uparrow}
            (W_{\downarrow}-V_{\downarrow}U_{\uparrow}) ] }  \ , \\   
\label{tdown}
t^M\sb{\uparrow\downarrow}   & = & \frac{1}
      {1-F\sb{\downarrow}U\sb{\downarrow}-F\sb{\uparrow}
      [V\sb{\uparrow}+F\sb{\downarrow}
         (W\sb{\uparrow}-V\sb{\uparrow}U\sb{\downarrow})]} 
       ( -IS\sb{-})  \     \nonumber   \\ 
   & = & ( - IS_{-})  \frac{1}
      {1-F\sb{\uparrow}U\sb{\uparrow}-F\sb{\downarrow}
       [V\sb{\downarrow}+F\sb{\uparrow}
            (W\sb{\downarrow}-V\sb{\downarrow}U\sb{\uparrow})]}  \  ,  
   \nonumber  \\  
  \\
t^M\sb{\downarrow\uparrow} & = & \frac{1}
      {1-F\sb{\uparrow}U\sb{\uparrow}-F\sb{\downarrow}
       [V\sb{\downarrow}+F\sb{\uparrow}
            (W\sb{\downarrow}-V\sb{\downarrow}U\sb{\uparrow})]} 
       ( - IS\sb{+}) \    \nonumber   \\
                           & = & ( - IS\sb{+})    
      \frac{1} 
      {1-F\sb{\downarrow}U\sb{\downarrow}-F\sb{\uparrow}
       [V\sb{\uparrow}+F\sb{\downarrow}
            (W\sb{\uparrow}-V\sb{\uparrow}U\sb{\downarrow})]}   \ .   \nonumber  \\
\end{eqnarray}
\end{subequations}
$V_{\uparrow}$\ $(V_{\downarrow})$ is the spin-diagonal component
of the interaction between 
a carrier with $\uparrow (\downarrow)$ spin
and the local potential on the \textit{M} ion
 embedded in the medium of $\Sigma_{\uparrow}$ ($\Sigma_{\downarrow}$).
A carrier with $\uparrow (\downarrow)$ spin which has already flipped 
in the previous scattering is subjected to $U_{\uparrow}$\ $(U_{\downarrow})$ on the \textit{M} ion
 embedded in the medium,
wherein the \textit{d} spin operator $S_z$ is replaced by $S_z-1$\ ($S_z+1$).
Furthermore, $W_{\uparrow}$ ($W_{\downarrow }$)
 represents the interaction energy required by a carrier
with $\uparrow$\ ($\downarrow$) spin to flip and then reverse its spin.
\section{Dynamical CPA --- Locator formalism}
\subsection{CPA locator condition}
In this subsection, we briefly outline an alternative but 
equivalent condition of the CPA,
which is an extension of the CPA using the locator formalism \cite{Kubo74}.
Assuming that the spin-dependent effective medium 
surrounds an arbitrary site \textit{n},
 we consider the transfer of the carrier with spin $\mu $ between site \textit{n} and the effective medium
 (i.e., $\Sigma _{\uparrow }$ and $ \Sigma _{\downarrow }$)
by the site-renormalized interactor $J_{\mu }$. 
Then, the propagator $G^A\ (G^M)$ associated with the real potential of $u_n^A\ (u_n^M)$
embedded at site \textit{n} in the medium is defined by
\begin{eqnarray}
G^A &=&  \frac{1}{\omega -  u^A_n - \sum_{\mu } J_{\mu } a^{\dagger}_{n \mu }a_{n \mu } } \ , \\
G^M &=& \frac{1}{\omega -  u^M_n - \sum_{\mu } J_{\mu } a^{\dagger}_{n \mu }a_{n \mu } } \ . 
\end{eqnarray} 
When we set the coherent potential $\Sigma_{\mu } $ on the site \textit{n} in the effective medium,
the reference propagator,
\begin{eqnarray}
P &=&  \frac{1}
{\omega - \sum_{\mu } \Sigma_{\mu } a^{\dagger}_{n \mu }a_{n \mu } 
 - \sum_{\mu } J_{\mu } a^{\dagger}_{n \mu }a_{n \mu } } \ ,
\end{eqnarray} 
 is equivalent to the Green function for the effective medium.
Thus, the diagonal matrix element of $P$ is 
equal to $F_{\mu }$ defined by Eq.\ (\ref{F}):
\begin{eqnarray}
F_{\mu }(\omega) &=& \langle n \mu | P | n \mu \rangle =
 \frac{1}{\omega - \Sigma_{\mu } - J_{\mu }} \ .
\label{JJ}
\end{eqnarray}
Equation (\ref{JJ}) gives the relationship between
 $J_{\mu }$ and $F_{\mu }$;
 ${\cal L}_{\mu } \equiv 1/(\omega -\Sigma_{\mu } )$ is called a locator.
Hereafter, for the sake of simplicity, the site-diagonal elements in the Wannier representation
$\langle n \mu | G^A | n \nu  \rangle $
 are written as $G^A_{\mu \nu } $ ($\mu, \nu =\uparrow, $ or $\downarrow $). 
Then, the spin-diagonal element of $ G^A$ is given by
\begin{eqnarray}
F_{\mu }^A(\omega ) = G^A_{\mu \mu } & = &  \frac{1}{\omega -  E_A -  J_{\mu } } \ ,
\label{FA} 
\end{eqnarray} 
and the spin-off-diagonal elements are $G^A_{\uparrow \downarrow }=G^A_{\downarrow \uparrow }=0$.
The site-diagonal elements of $G^M$ are obtained after a somewhat complicated calculation 
using the commutation relationships between the components of ${\bf S}$
but with no further approximations, as
\begin{widetext}
\begin{subequations}
\begin{eqnarray}
G^M_{\uparrow \uparrow } &=& \frac{\omega -E_M-I(S_z+1)-J_{\downarrow }}
   {[\omega -(E_M-IS_z)-J_{\uparrow }][\omega -E_M-I(S_z+1)-J_{\downarrow }]
    -I^2[S(S+1)-S^2_z-S_z]} \\
G^M_{\downarrow \downarrow } &=& \frac{\omega -E_M+I(S_z-1)-J_{\uparrow }}
   {[\omega -(E_M+IS_z)-J_{\downarrow }][\omega -E_M+I(S_z-1)-J_{\uparrow }]
   -I^2[S(S+1)-S^2_z+S_z]}  \\
G^M_{\uparrow \downarrow } &=& \frac{1}
   {[\omega -(E_M-IS_z)-J_{\uparrow }][\omega -E_M-I(S_z+1)-J_{\downarrow }]
    -I^2[S(S+1)-S^2_z-S_z]} (-IS_{-}) \nonumber \\
\\
G^M_{\downarrow \uparrow  } &=& \frac{1}
   {[\omega -(E_M+IS_z)-J_{\downarrow }][\omega -E_M+I(S_z-1)-J_{\uparrow }]
   -I^2[S(S+1)-S^2_z+S_z]} (-IS_{+}) \nonumber \ . \\
\end{eqnarray}
\end{subequations}
\end{widetext}
Note that the site-diagonal elements of $G^M$ involve spin operators.
Thus, $F^M_\mu (\omega ) $ is defined as the thermal average of 
the spin-diagonal element $G^M_{\mu \mu }$ by
\begin{equation}
F^M_{\mu} (\omega ) = \langle G^M_{\mu \mu} \rangle
 = \sum_{S_z=-S}^S G^M_{\mu \mu}(S_z)
{\rm exp} (\lambda S_z)
/\sum_{S_z=-S}^S {\rm exp} (\lambda S_z) \ ,
\label{FM}
\end{equation}
where $\lambda\ (\equiv h/k_BT) $ is determined so as to reproduce a given
value of $\langle S_z \rangle $ [see Eq.\ (\ref{THREM})].
Note that the spin-off-diagonal elements $\langle G^M_{\uparrow \downarrow } \rangle
 = \langle G^M_{\downarrow \uparrow } \rangle = 0$,
because $G^M_{\uparrow \downarrow } $ ($G^M_{\downarrow \uparrow } $) 
includes $S_{-}$ ($S_{+}$) in their final form.
Finally, the CPA condition in the locator formula is given by
\begin{eqnarray}
F_{\mu }(\omega ) & = &  (1-x) F^A_{\mu }(\omega ) + x F^M_{\mu }(\omega ) .
\label{LocatorCPA}
\end{eqnarray} 
When $F_{\mu }$ is given,
$J_{\mu }$ is calculated by Eqs. (\ref{F}) and (\ref{JJ})
[i.e., $J_{\mu }=(\Delta^2 /4 )F_{\mu }$ for the model band defined by Eq.\ (\ref{rho})].
Then, $F^A_{\mu }$ and  $F^M_{\mu }$ are calculated by Eqs.\ (\ref{FA}) and (\ref{FM}), 
and consequently,
$F_{\mu }$ is again obtained by Eq.\ (\ref{LocatorCPA}).
Therefore, $F_{\mu }$ and $J_{\mu }$ are determined self-consistently.

\subsection{Local DOS}
The advantage of the locator formula CPA is that 
it is straightforward to determine species-resolved DOS, i.e., DOS 
associated with each kind of ion in the alloy.
The local density of states (local DOS) at the $A\ (M)$ site is obtained by
\begin{subequations}
\begin{eqnarray}
D^A_{\mu } (\omega ) &=& - \frac{1}{\pi} {\rm Im} F^A_{\mu } (\omega ) \ , \\
D^M_{\mu } (\omega ) &=& - \frac{1}{\pi} {\rm Im} F^M_{\mu } (\omega ) \ .
\end{eqnarray}
\end{subequations}
In actual calculations, first we determined $F_{\mu }$ and $\Sigma _{\mu }$
 by the \textit{t} matrix formula CPA, and then
calculated $J_{\mu }$, and consequently $D^A_{\mu } (\omega ) $ and $D^M_{\mu } (\omega )$.
We numerically verified the relations
\begin{eqnarray}
\int ^{\infty}_{-\infty} D^A_{\mu }(\omega ) d \omega 
=  \int ^{\infty}_{-\infty} D^M_{\mu }(\omega ) d \omega 
= 1 \ ,
\end{eqnarray} 
and 
\begin{eqnarray}
D_{\mu }(\omega )  = (1-x) D^A_{\mu }(\omega ) + x D^M_{\mu }(\omega ) \ ,
\end{eqnarray} 
which are a consequence of the CPA locator condition, Eq.\ (\ref{LocatorCPA}).

\subsection{Spin-coupling strength $Q(\omega )$}
We define the spin-coupling strength $Q(\omega )$
as the normalized (negative) inner product 
between the carrier spin and localized spin on the \textit{M} site:
where $D^M(\omega ) = D^M_{\uparrow } (\omega ) + D^M_{\downarrow }(\omega ) $ and 
\begin{widetext}
\begin{eqnarray}
C(\omega ) & \equiv &  \sum_{\mu \nu } \langle \delta (\omega -H) \ a^{\dagger}_{n \mu }
\sigma_{\mu \nu } \cdot \mathbf{S} a_{n \nu } \rangle \nonumber \\
&=& - \frac{1}{\pi} \left\langle \textrm{Im} \sum_{\mu \nu } \langle n \nu | G^M(\omega ) |n\mu \rangle
\langle n \mu | a^{\dagger}_{n \mu }
\sigma_{\mu \nu } \cdot \mathbf{S} a_{n \nu } | n \nu \rangle \right\rangle \nonumber \\
&=& - \frac{1}{\pi} \textrm{Im} \left\langle (G^M_{\uparrow \uparrow }-G^M_{\downarrow \downarrow })
S_z +G^M_{\uparrow \downarrow }S_{+} +G^M_{\downarrow \uparrow }S_{-} \right\rangle \ .
\end{eqnarray} 
\end{widetext}

\end{document}